\documentclass[5p, sort&compress]{elsarticle}
\usepackage{graphicx}
\usepackage{amsmath}
\usepackage{amssymb}
\usepackage{color}
\usepackage{bm}
\usepackage{hyperref}
\usepackage{natbib}
\hypersetup{
  pdfnewwindow=true, colorlinks=true,
  linkcolor=blue, anchorcolor=blue,
  citecolor=blue, filecolor=blue,
  menucolor=blue, urlcolor=blue}
\graphicspath{ {./}{./figures/} }
\journal{Computer Physics Communications}
\begin{document}
\begin{frontmatter}
\title{{\sc Perturbo}: a software package for \textit{ab initio} electron-phonon interactions, \\ charge transport and ultrafast dynamics}
\author[CIT]{Jin-Jian Zhou\corref{cor1}}
\author[CIT]{Jinsoo Park}
\author[CIT]{I-Te Lu}
\author[CIT]{Ivan Maliyov}
\author[CIT]{Xiao Tong}
\author[CIT]{Marco Bernardi\corref{cor1}}
\cortext[cor1]{Corresponding author. \\ \textit{E-mail addresses:}  
 \\  jjzhou@caltech.edu (J.-J. Zhou),  bmarco@caltech.edu (M. Bernardi).}

\address[CIT]{Department of Applied Physics and Materials Science, California Institute of Technology, Pasadena, CA 91125, USA.}
\begin{abstract}
{\sc Perturbo} is a software package for first-principles calculations of charge transport and ultrafast carrier dynamics in materials. The current version focuses on electron-phonon interactions and can compute phonon-limited transport properties such as the conductivity, carrier mobility and Seebeck coefficient. It can also simulate the ultrafast nonequilibrium electron dynamics in the presence of electron-phonon scattering. {\sc Perturbo} uses results from density functional theory and density functional perturbation theory calculations as input, and employs Wannier interpolation to reduce the computational cost. It supports norm-conserving and ultrasoft pseudopotentials, spin-orbit coupling, and polar electron-phonon corrections for bulk and 2D materials. Hybrid MPI plus OpenMP parallelization is implemented to enable efficient calculations on large systems (up to at least 50 atoms) using high-performance computing. Taken together, {\sc Perturbo} provides efficient and broadly applicable \textit{ab initio} tools to investigate electron-phonon interactions and carrier dynamics quantitatively in metals, semiconductors, insulators, and 2D materials. 
\end{abstract}
%
%% keyword
\end{frontmatter}

\begin{small}
\noindent
{\bf PROGRAM SUMMARY} \\
  %Delete as appropriate.

\noindent
{\em Program Title: } {\sc Perturbo} \\
{\em Program summary URL:} \url{https://perturbo-code.github.io} \\ 
{\em Licensing provisions:} GNU General Public Licence 3.0 \\
{\em Programming language:}  {\sc Fortran}, {\sc Python} \\
{\em External routines/libraries:}  {\sc LAPACK}, {\sc HDF5}, {\sc MPI}, {\sc OpenMP}, FFTW, {\sc Quantum-ESPRESSO} \\
{\em Nature of problem:} 
%Describe the nature of the problem here, (approx. 50-250 words) \\
Computing  transport properties from first-principles in materials, including the electrical conductivity, carrier mobility and Seebeck coefficient; Simulating ultrafast nonequilibrium electron dynamics, such as the relaxation of excited carriers via interactions with phonons. \\
{\em Solution method:} 
 %Describe the method solution here, (approx. 50-250 words)
We implement the first-principles Boltzmann transport equation, which employs materials properties such as the electronic structure, lattice dynamics, and electron-phonon collision terms computed with density functional theory and density functional perturbation theory. The Boltzmann transport equation is solved numerically to compute charge transport and simulate ultrafast carrier dynamics. Wannier interpolation is employed to reduce the computational cost. \\
{\em Additional comments:} 
Hybrid MPI plus OpenMP parallelization is implemented to run large calculations and take advantage of high-performance computing. Most results are output to HDF5 file format, which is portable and convenient for post-processing using high-level languages such as Python and Julia. 
\end{small}
%\maketitle
%introduction
\section{Introduction}
Understanding the dynamical processes involving electrons, lattice vibrations (phonons), atomic defects, and photons in the solid state is key to developing the next generation of materials and devices~\cite{Rossi2002,Ulbricht2011}. 
Due to the increasing complexity of functional materials, there is a critical need for computational tools that can take into account the atomic and electronic structure of materials and make quantitative predictions on their physical properties.
%
%accurate calculations of electron interactions and charge carrier dynamics~\cite{Egger2018}. Computational methods able to take into account the material atomic and electronic structures to make accurate predictions are thus poised to advance computational materials physics.
%
The vision behind {\sc Perturbo} is to provide a unified platform and a validated code that can be applied broadly to compute the interactions, transport and ultrafast dynamics of electrons and excited states in materials~\cite{Egger2018}.
The goal is to facilitate basic scientific discoveries in materials and devices by advancing microscopic understanding of carrier dynamics, while creating a sustainable software element able to address the demands of the computational physics community. \\
%
%fundamental needs in energy, science and technology, and industry research.
%two-fold $-$ facilitating basic scientific discoveries in materials and devices by advancing microscopic understanding of carrier dynamics, and creating a sustainable software element able to address fundamental needs in energy, science and technology, and industry research. 
%
\indent
{\sc Perturbo} builds on established first-principles methods. It uses density functional theory (DFT) and density functional perturbation theory (DFPT)~\cite{Baroni2001} as a starting point for computing electron dynamics.  
It reads the output of DFT and DFPT calculations, for now from the {\sc Quantum Espresso} (QE) code~\cite{Giannozzi2009,Giannozzi2017}, and uses this data to compute electron interactions, charge transport and ultrafast dynamics. 
The current distribution focuses on electron-phonon ($e$-ph) interactions and the related phonon-limited transport properties~\cite{Ziman2007}, including the electrical conductivity, mobility and the Seebeck coefficient. 
It can also simulate the ultrafast nonequilibrium electron dynamics in the presence of $e$-ph interactions.
The developer branch, which is not publicly available yet, also includes routines for computing spin~\cite{Park2020}, electron-defect~\cite{Lu2019,Lu2019a}, and electron-photon interactions~\cite{Chen2019}, as well as advanced methods to compute the ultrafast dynamics of electrons and phonons in the presence of electric and magnetic fields. These additional features will be made available in future releases.
\\
\indent
The transport module of {\sc Perturbo} enables accurate calculations of charge transport in a wide range of functional materials. In its most basic workflow, {\sc Perturbo} computes the conductivity and mobility as a function of temperature and carrier concentration, either within the relaxation time approximation (RTA) or with an iterative solution of the linearized Boltzmann transport equation (BTE)~\cite{Zhou2016,Zhou2018}. 
The ultrafast dynamics module explicitly evolves in time the electron BTE (while keeping the phonon occupations fixed), enabling investigations of the ultrafast electron dynamics starting from a given initial electron distribution~\cite{Jhalani2017}. 
Our routines can carry out these calculations in metals, semiconductors, insulators, and 2D materials. 
An efficient implementation of long-range $e$-ph interactions is employed for polar bulk and 2D materials.
%such as III$-$V and II$-$VI semiconductors, alkali halides, and oxides. 
%, including polar materials (e.g., III$-$V and II$-$VI semiconductors, alkali halides, and oxides) thanks to an efficient implementation of long-range $e$-ph interactions~\cite{Zhou2016}, and 2D materials. 
Materials with spin-orbit coupling (SOC) are treated using fully relativistic pseudopotentials~\cite{Park2020,Zhou2018}. Both norm-conserving and ultrasoft pseudopotentials are supported. Quantities related to $e$-ph interactions can be easily obtained, stored and analyzed. \\
\indent
{\sc Perturbo} is implemented in modern Fortran with a modular code design. All calculations employ intuitive workflows. The code is highly efficient thanks to its hybrid MPI (Message Passing Interface) and OpenMP (Open Multi-Processing) parallelization. It can run on record-large unit cells with up to at least 50 atoms~\cite{Lee2018}, and its performance scales up to thousands of CPU cores. 
It conveniently writes files using the HDF5 format, and is suitable for both high-performance supercomputers and smaller computer clusters. \\
%In benchmark calculations, it is orders of magnitude faster than existing codes for computing $e$-ph interactions and transport; 
%It can run on record-large unit cells with up to 50$-$100 atoms~\cite{Lee2018}, and its performance scales up to thousands of CPU cores. \\
%
%The code can carry out many workflows (e.g., ultrafast carrier dynamics and transport in 2D materials, among others) that are not possible with other existing codes.
%
\indent
Target users include both experts in first-principles calculations and materials theory as well as experimental researchers and teams in academic or national laboratories investigating charge transport, ultrafast spectroscopy, advanced functional materials, and semiconductor or solid-state devices.  {\sc Perturbo} will equip these users with an efficient quantitative tool to investigate electron interactions and dynamics in broad families of materials, filling a major void in the current software ecosystem. \\
%
%allow these users to quantitatively study charge, spin, and energy transport, as well as ultrafast electron and excited state dynamics, in broad families of novel functional materials, filling a major void in the current software ecosystem.
%
\indent
 The paper is organized as follows: Sec.~\ref{sec:method} discusses the theory and numerical methods implemented in the code; Sec.~\ref{sec:cap} describes the code capabilities and workflows; Sec.~\ref{sec:tech} delves deeper into selected technical aspects; Sec.~\ref{sec:example} shows several example calculations provided as tutorials in the code; Sec.~\ref{sec:parallel} discusses the parallelization strategy and the scaling of the code on high performance supercomputers. We conclude in Sec.~\ref{sec:conc} by summarizing the main features and planned future development of {\sc Perturbo}.

\section{Methodology} \label{sec:method}
% general introduction of methods and formula.
\subsection{Boltzmann transport equation}
The current release of {\sc Perturbo} can compute charge transport and ultrafast dynamics in the framework of the semiclassical BTE. The BTE describes the flow of the electron occupations $f_{n\bm{k}}(\bm{r},t)$ in the phase-space variables of relevance in a periodic system, the crystal momentum $\bm{k}$ and spatial coordinate $\bm{r}$:
\begin{equation}
\begin{aligned} 
\frac{\partial f_{n \bm{k}}(\bm{r}, t)}{\partial t}   =  &  -\left[\nabla_{\bm{r}} f_{n \bm{k}}(\bm{r}, t) \cdot \bm{v}_{n \bm{k}} + \hbar^{-1} \nabla_{\bm{k}} f_{n \bm{k}}(\bm{r}, t) \cdot \bm{F}\right] 
\\ &+ \mathcal{I}\left[f_{n \bm{k}}\right],
 \end{aligned}
 \label{eq:BTE_full}
\end{equation}
where $n$ is the band index and $v_{n\bm{k}}$ are band velocities. The time evolution of the electron occupations is governed by the so-called drift term due to external fields $\bm{F}$ and the collision term $\mathcal{I}\left[f_{n \bm{k}}\right]$, which captures electron scattering processes due to phonons or other mechanisms~\cite{Bernardi2016}. In {\sc Perturbo}, the fields are assumed to be slowly varying and the material homogeneous, so $f_{n \bm{k}}$ does not depend on the spatial coordinates and its spatial dependence is not computed explicitly.\\
\indent
The collision integral $\mathcal{I}\left[f_{n \bm{k}}\right]$ is a sum over a large number of scattering processes in momentum space, and it is very computationally expensive because it involves Brillouin zone (BZ) integrals on fine grids. Most analytical and computational treatments simplify the scattering integral with various approximations. A common one is the RTA, which assumes that the scattering integral is proportional to the deviation $\delta f_{n \bm{k}}$ of the electron occupations from the equilibrium Fermi-Dirac distribution, $\mathcal{I}\left[f_{n \bm{k}}\right] = -\delta f_{n \bm{k}} / \tau$; the relaxation time $\tau$ is either treated as a constant empirical parameter~\cite{Madsen2006,Pizzi2014} or as a state-dependent quantity, $\tau_{n \bm{k}}$. \\
\indent
{\sc Perturbo} implements the first-principles formalism of the BTE, which employs materials properties obtained with quantum mechanical approaches, using the atomic structure of the material as the only input.  The electronic structure is computed using DFT and the lattice dynamical properties using DFPT. The scattering integral is computed only for $e$-ph processes in the current release, while other scattering mechanisms, such as electron-defect and electron-electron scattering, are left for future releases. 
The scattering integral due to $e$-ph processes can be written as 
\begin{equation}
\begin{aligned}
\mathcal{I}^{e\mathrm{-ph}}\left[f_{n \bm{k}}\right]  = &
-\frac{2 \pi}{\hbar} \frac{1}{{\mathcal{N}_{\bm{q}}}} \sum_{m \bm{q} \nu}\left|g_{m n \nu}(\bm{k}, \bm{q})\right|^{2} \\ &  \times
 [\delta\left(\varepsilon_{n \bm{k}}-\hbar \omega_{\nu \bm{q}}-\varepsilon_{m \bm{k}+\bm{q}}\right) \times F_{\mathrm{em}}  \\ & +\delta\left(\varepsilon_{n \bm{k}}+\hbar \omega_{\nu \bm{q}}-\varepsilon_{m \bm{k}+\bm{q}}\right) \times F_{\mathrm{abs}} ],
\end{aligned}
\label{eq:e_ph_scattering}
\end{equation}
where $\mathcal{N}_{\bm{q}}$ is the number of $\bm{q}$-points used in the summation, and $g_{m n  \nu}(\bm{k}, \bm{q})$ are $e$-ph matrix elements (see Sec.~\ref{sec:wan_eph}) quantifying the probability amplitude for an electron to scatter from an initial state $\left|n\bm{k}\right\rangle$ to a final state $\left|m{\bm{k}+\bm{q}}\right\rangle$, by emitting or absorbing a phonon with wavevector $\bm{q}$ and mode index $\nu$; here and below, $\varepsilon_{n\bm{k}}$ and $\hbar\omega_{\nu \bm{q}}$ are the energies of electron quasiparticles and phonons, respectively. The phonon absorption ($F_{\mathrm{abs}}$) and emission ($F_{\mathrm{em}}$) terms are defined as
\begin{equation} 
\begin{aligned}
F_{\mathrm{abs}}  &= f_{n \bm{k}}\left(1-f_{m \bm{k}+\bm{q}}\right) N_{\nu \bm{q}}-f_{m \bm{k}+\bm{q}}\left(1-f_{n \bm{k}}\right)\left(N_{\nu \bm{q}}+1\right), \\
F_{\mathrm{em}}  &=  f_{n \bm{k}}\left(1-f_{m \bm{k}+\bm{q}}\right)\left(N_{\nu \bm{q}}+1\right)-f_{m \bm{k}+\bm{q}}\left(1-f_{n \bm{k}}\right) N_{\nu \bm{q}}
\end{aligned}
\end{equation}
where $N_{\nu \bm{q}}$ are phonon occupations. 
%This is true only for steady-state (not ultrafast): In this framework, the temperature dependence of the collision processes is due to the electron and phonon occupations, $f_{n\bm{k}}(T)$ and $N_{\nu \bm{q}} (T)$, which 
%
\subsubsection{Ultrafast carrier dynamics}
{\sc Perturbo} can solve the BTE numerically to simulate the evolution in time $t$ of the electron occupations $f_{n\bm{k}}(t)$ due to $e$-ph scattering processes, starting from an initial nonequilibrium distribution. In the presence of a slowly varying external electric field, and assuming the material is homogeneous, we rewrite the BTE in Eq.~(\ref{eq:BTE_full}) as 
\begin{equation}
 \frac{\partial f_{n \bm{k}}(t)}{\partial t} = \frac{e\bm{E}}{\hbar}  \cdot  \nabla_{\bm{k}} f_{n \bm{k}}(t) +\mathcal{I}^{e\mathrm{-ph}}\left[f_{n \bm{k}}\right],
 \label{eq:BTE_E}
\end{equation}
% *** WE SHOULD SWITCH TO e > 0 rather than $e = - |e|$
where $e$ is the electron charge and $\bm{E}$ the external electric field.  
We solve this non-linear integro-differential equation numerically (in the current release, only for $\bm{E}=0$), using explicit time-stepping with the 4th-order Runge-Kutta (RK4) or Euler methods. The RK4 solver is used by default due to its superior accuracy. The Euler method is much faster than RK4, but it is only first-order accurate in time, so it should be tested carefully and compared against RK4. \\
\indent
Starting from an initial nonequilibrium electron distribution $f_{n\bm{k}}(t_{0})$, we evolve in time Eq.~(\ref{eq:BTE_E}) with a small time step (typically in the order of 1 fs) to obtain $f_{n\bm{k}}(t)$ as a function of time. 
One application of this framework is to simulate the equilibration of excited electrons~\cite{Jhalani2017,Bernardi2014}, in which we set $\bm{E}=0$ and simulate the time evolution of the excited electron distribution as it approaches its equilibrium Fermi-Dirac value through $e$-ph scattering processes. 
The phonon occupations are kept fixed (usually, to their equilibrium value at a given temperature) in the current version of the code. Another application, which is currently under development, is charge transport in high electric fields. 
\subsubsection{Charge transport}
In an external electric field, the drift and collision terms in the BTE balance out at long enough times $-$ the field drives the electron distribution out of equilibrium, while the collisions tend to restore equilibrium. At steady state, a nonequilibrium electron distribution is reached, for which $\partial f_{n\bm{k}} / \partial t = 0$. The BTE for transport at steady state becomes
\begin{equation}
 -\frac{e\bm{E}}{\hbar}  \cdot  \nabla_{\bm{k}} f_{n \bm{k}}(t) = \mathcal{I}^{e\mathrm{-ph}}\left[f_{n \bm{k}}\right].
 \label{eq:BTE_T}
\end{equation}
When the electric field is relatively weak, the steady-state electron distribution deviates only slightly from its equilibrium value. As is usual, we expand $f_{n\bm{k}}$ around the equilibrium Fermi-Dirac distribution, $f_{n\bm{k}}^{0}$, and keep only terms linear in the electric field:
\begin{equation}
\begin{aligned} 
f_{n\bm{k}} &=f_{n\bm{k}}^{0}+f_{n\bm{k}}^{1}+\mathcal{O}\left(E^{2}\right) \\ & = f_{n\bm{k}}^{0} + e \bm{E} \cdot \bm{F}_{n\bm{k}} \frac{\partial f_{n\bm{k}}^{0}}{\partial \varepsilon_{n\bm{k}}}+\mathcal{O}\left(E^{2}\right),
\end{aligned}
\label{eq:dist_exp}
\end{equation}
where $\bm{F}_{n\bm{k}}$ characterizes the first-order deviation from equilibrium of the electron distribution. We substitute Eq.~(\ref{eq:dist_exp}) into both sides of Eq.~(\ref{eq:BTE_T}), and obtain a linearized BTE 
for the distribution deviation $\bm{F}_{n\bm{k}}$ keeping only terms up to first-order in the electric field:
\begin{equation}
\bm{F}_{n\bm{k}} = \tau_{n\bm{k}} \bm{v}_{n\bm{k}} + 
 \frac{\tau_{n\bm{k}}}{\mathcal{N}_{\bm{q}}} \sum_{m,\nu\bm{q}} \bm{F}_{m\bm{k}+\bm{q}} W_{n\bm{k}, m\bm{k}+\bm{q}}^{\nu \bm{q}}\,,
\label{eq:BTE_lin}
\end{equation}
where $\tau_{n\bm{k}}$ is the electron relaxation time, computed as the inverse of the scattering rate, $\tau_{n\bm{k}} = \Gamma_{n\bm{k}}^{-1}$. The scattering rate $\Gamma_{n\bm{k}}$ is given by
\begin{equation}
\Gamma_{n\bm{k}} = \frac{1}{\mathcal{N}_{\bm{q}}}\sum_{m,\nu\bm{q}} W_{n\bm{k}, m\bm{k}+\bm{q}}^{\nu \bm{q}}.
\label{eq:scatter_rate}
\end{equation}
The scattering probability $W_{n\bm{k}, m\bm{k}+\bm{q}}^{\nu \bm{q}}$ involves both phonon emission and absorption processes:
\begin{equation}
\begin{aligned} 
W_{n\bm{k}, m\bm{k}+\bm{q}}^{\nu \bm{q}}  & =  \frac{2\pi}{\hbar}\left|g_{m n \nu}(\bm{k}, \bm{q}) \right|^{2} \\
& \times [  \delta\left(\varepsilon_{n \bm{k}}-\hbar\omega_{\nu \bm{q}}-\varepsilon_{m \bm{k}+\bm{q}}\right)
\left(1+N^{0}_{\nu \bm{q}}-f^{0}_{m \bm{k}+\bm{q}}\right)  \\ 
 & + \delta\left(\varepsilon_{n \bm{k}}+\hbar\omega_{\nu \bm{q}}-\varepsilon_{m \bm{k}+\bm{q}}\right)
  \left(N^{0}_{\nu \bm{q}}+f^{0}_{m \bm{k}+\bm{q}}\right) ].
 \end{aligned}
 \label{eq:BTE_W}
\end{equation}
where $N^{0}_{\nu \bm{q}}$ are equilibrium Bose-Einstein phonon occupations. Note that since $\tau_{n\bm{k}}$ is an electron quasiparticle lifetime, it can be written equivalently as the imaginary part of the $e$-ph self-energy~\cite{Mahan2000}, $\tau_{n\bm{k}}^{-1} = 2\mathrm{Im}\Sigma_{n\bm{k}}^{e-\mathrm{ph}}/\hbar$. \\
\indent
In the RTA, we neglect the second term in Eq.~(\ref{eq:BTE_lin}) and obtain $\bm{F}_{nk}=\tau_{n\bm{k}} \bm{v}_{n\bm{k}}$. In some cases, the second term in Eq.~(\ref{eq:BTE_lin}) cannot be neglected. In metals, a commonly used scheme to approximate this term is to add a factor of 
$(1 - \mathrm{cos}\,\theta_{\bm{k},\bm{k}+\bm{q}})$ to Eq.~(\ref{eq:BTE_W}), where $\theta_{\bm{k},\bm{k}+\bm{q}}$ is the scattering angle between $\bm{k}$ and $\bm{k}+\bm{q}$. The resulting so-called  ``transport relaxation time''  is then used to compute the transport properties~\cite{Mahan2000}.
{\sc Perturbo} implements a more rigorous approach and directly solves Eq.~(\ref{eq:BTE_lin}) using an iterative method~\cite{Li2015,Fiorentini2016}, for which we rewrite Eq.~(\ref{eq:BTE_lin}) as 
\begin{equation}
\bm{F}_{n\bm{k}}^{i+1}=  \bm{F}_{nk}^{0} + 
 \frac{\tau_{n\bm{k}}}{\mathcal{N}_{\bm{q}}} \sum_{m,\nu\bm{q}} \bm{F}_{m\bm{k}+\bm{q}}^{i} W_{n\bm{k}, m\bm{k}+\bm{q}}^{\nu \bm{q}}\,.
 \label{eq:BTE_iter}
\end{equation}
In the iterative algorithm, we choose in the first step  $\bm{F}_{nk}^{0}=\tau_{n\bm{k}} \bm{v}_{n\bm{k}}$, and then compute the following steps using Eq.~(\ref{eq:BTE_iter}) until the difference $|\bm{F}_{n\bm{k}}^{i+1} - \bm{F}_{n\bm{k}}^{i}|$ is within the convergence threshold.\\
\indent
Once $\bm{F}_{nk}$ has been computed, either within the RTA or with the iterative solution of the BTE in Eq.~(\ref{eq:BTE_iter}), the conductivity tensor is obtained as 
\begin{equation}
\begin{aligned}
\sigma_{\alpha\beta} &= \frac{1}{\Omega \bm{E}_{\beta}}\cdot\frac{S}{\mathcal{N}_{\bm{k}}}\sum_{n\bm{k}}-e \bm{v}_{n\bm{k}}^{\alpha}\cdot f^{1}_{n\bm{k}} \\
 & = \frac{e^{2}S}{\mathcal{N}_{\bm{k}} \Omega}\sum_{n\bm{k}}\bm{v}_{n\bm{k}}^{\alpha} \bm{F}_{n\bm{k}}^{\beta} \left(-\frac{\partial f_{n\bm{k}}^{0}}{\partial\varepsilon_{n\bm{k}}}\right),
\end{aligned}
 \label{eq:cond}
\end{equation}
where $\alpha$ and $\beta$ are Cartesian directions, $\Omega$ is the volume of the unit cell, and $S$ is the spin degeneracy. We also compute the carrier mobility tensor, $\mu_{\alpha \beta} = \sigma_{\alpha\beta} / (e\,n_{c})$, by dividing the conductivity tensor through the carrier concentration $n_c$. \\
\indent
In our implementation, we conveniently rewrite Eq.~(\ref{eq:cond}) as 
\begin{equation}
\sigma_{\alpha\beta} = e^{2} \int dE (-\partial f^{0}/\partial E) \Sigma_{\alpha\beta} (E),
\label{eq:cond_tdf}
\end{equation}
where $\Sigma_{\alpha\beta}(E)$ is the transport distribution function (TDF) at energy $E$, 
\begin{equation}
\Sigma_{\alpha\beta} (E) = \frac{S}{\mathcal{N}_{\bm{k}} \Omega} \sum_{n\bm{k}} \bm{v}_{n\bm{k}}^{\alpha} \bm{F}_{nk}^{\beta} \delta(E-\varepsilon_{n\bm{k}}),
\label{eq:cond_tdf2}
\end{equation}
which is computed in {\sc Perturbo} using the tetrahedron integration method~\cite{Bloechl1994}.
The integrand in Eq.~(\ref{eq:cond_tdf}) can be used to characterize the contributions to transport 
as a function of electron energy~\cite{Zhou2016,Zhou2018}. The code can also compute the Seebeck coefficient $\bm{S}$ from the TDF, using 
\begin{equation}
\left[\bm{\sigma}\bm{S}\right]_{\alpha\beta} = \frac{e}{T} \int dE (-\partial f^{0}/\partial E) (E - \mu) \Sigma_{\alpha\beta} (E),
\label{eq:seebeck}
\end{equation}
where $\mu$ is the chemical potential and $T$ is the temperature. 
\subsection{Electrons, phonons, and $e$-ph interactions}
As discussed above, the electronic structure and phonon dispersion are computed with DFT and DFPT. In principle, the $e$-ph matrix elements can also be computed with these methods and used directly for carrier dynamics calculations. However, to converge transport and ultrafast dynamics with the BTE, the $e$-ph matrix elements and the scattering processes need to be computed on ultra-dense $\bm{k}$- and $\bm{q}$-point BZ grids with roughly $100\times 100 \times 100$ or more points. Therefore, the computational cost is prohibitive for direct DFPT calculations, and we resort to interpolation techniques to obtain the $e$-ph matrix elements and other relevant quantities on fine grids, starting from DFT and DFPT calculations on coarser BZ grids, typically of order $10\times 10 \times 10$. 
\subsubsection{Wannier interpolation of the electronic structure} \label{sec:wannier}
We use the Wannier interpolation to compute efficiently the electron energy and band velocity on ultra-fine $\bm{k}$-point grids~\cite{Marzari2012}. We first perform DFT calculations on a regular coarse grid with points $\bm{k}_{c}$, and obtain the electron energies $\varepsilon_{n\bm{k}_{c}}$ and Bloch wavefunctions $| \psi_{n\bm{k}_{c}} \rangle$.  We construct maximally localized Wannier functions $| n \bm{R}_{e} \rangle$, with index $n$ and centered in the cell at $\bm{R}_{e}$, from the Bloch wavefunctions using the {\sc Wannier90} code~\cite{Mostofi2014,Pizzi2020}: 
\begin{equation}
| n \bm{R}_{e} \rangle = \frac{1}{N_{e}} \sum_{m\bm{k}_{c}} e^{-i\bm{k}_{c} \cdot \bm{R}_{e} } 
\mathcal{U}_{mn}(\bm{k}_{c})\, | \psi_{m\bm{k}_{c}} \rangle ,
\label{eq:wannier}
\end{equation}
where $N_{e}$ is the number of $\bm{k}_{c}$-points in the coarse grid, and  $\mathcal{U}(\bm{k}_{c})$ are the unitary matrices transforming the Bloch wavefunctions to a Wannier gauge~\cite{Wang2006}, 
\begin{equation}
| \psi^{(W)}_{n\bm{k}_{c}} \rangle = \sum_{m} \mathcal{U}_{mn}(\bm{k}_{c})\,| \psi_{m\bm{k}_{c}} \rangle.
\label{eq:wannier_u}
\end{equation}
For entangled band structures, $\mathcal{U}(\bm{k}_{c})$ are not in general square matrices since they are also used to extract a smooth subspace from the original DFT Bloch eigenstates~\cite{Souza2001}. \\
\indent 
We compute the electron Hamiltonian in the Wannier function basis,
\begin{equation}
\begin{aligned}
H_{nn^{\prime}}(\bm{R}_{e}) &= \langle  n \bm{0} | \hat{H}  | n^{\prime}  \bm{R}_{e} \rangle \\
 &= \frac{1}{N_{e}} \sum_{\bm{k}_{c}} e^{-i\bm{k}_{c} \cdot \bm{R}_{e} } 
\left[ \mathcal{U}^{\dagger}(\bm{k}_{c}) H(\bm{k}_{c})\mathcal{U}(\bm{k}_{c}) \right]_{nn^{\prime}}\,
\end{aligned}
\label{eq:wannier_2}
\end{equation}
where $H(\bm{k}_{c})$ is the Hamiltonian in the DFT Bloch eigenstate basis, 
$H_{nm}(\bm{k}_{c}) = \varepsilon_{n\bm{k}_{c}} \delta_{nm}$. 
The Hamiltonian in the Wannier basis, $H_{nn^{\prime}}(\bm{R}_{e})$, can be seen as an \textit{ab initio} tight-binding model, with hopping integrals from the Wannier orbital $n^{\prime}$ in the cell at $\bm{R}_{e}$ to the Wannier orbital $n$ in the cell at the origin. Due to the localization of the Wannier orbitals,  the hopping integrals decay rapidly with $|\bm{R}_{e}|$, so a small set of  $\bm{R}_{e}$ vectors is sufficient to represent the electronic structure of the system. \\
\indent
Starting from $H_{nn^{\prime}}(\bm{R}_{e})$, we obtain the band energy $\varepsilon_{n\bm{k}}$ and band velocity $\bm{v}_{n\bm{k}}$ at any desired $\bm{k}$-point. We first compute the Hamiltonian matrix $H^{(W)}(\bm{k})$ in the basis of Bloch sums of Wannier functions using an inverse discrete Fourier transform, and then diagonalize it through a unitary rotation matrix $U(\bm{k})$ satisfying
\begin{equation}
H^{(W)}(\bm{k}) = 
\sum_{\bm{R}_{e}} e^{i \bm{k} \cdot \bm{R}_{e}} H(\bm{R}_{e}) 
= U(\bm{k}) H^{(H)}(\bm{k}) U^{\dagger}(\bm{k})\,,
\label{eq:wannier_3}
\end{equation}
where $H^{(H)}_{nm}(\bm{k}) = \varepsilon_{n\bm{k}}\delta_{nm}$,  and $\varepsilon_{n\bm{k}}$ and $U(\bm{k})$ are the eigenvalues and eigenvectors of $H^{(W)}(\bm{k})$, respectively. One can also obtain the corresponding interpolated Bloch eigenstates as
\begin{equation}
|\psi_{n\bm{k}}\rangle = \sum_{m} U_{mn}(\bm{k})\, |\psi^{(W)}_{m\bm{k}}\rangle
= \sum_{m} U_{mn}(\bm{k}) \sum_{\bm{R}_{e}} e^{i\bm{k}\cdot\bm{R}_{e}}   |m\bm{R}_{e}\rangle.
\label{eq:bloch_state}
\end{equation} 
The band velocity in the Cartesian direction $\alpha$ is computed as 
\begin{equation}
\hbar \bm{v}^{\alpha}_{n\bm{k}} = [U^{\dagger}(\bm{k}) H^{(W)}_{\alpha}(\bm{k}) U(\bm{k})]_{nn}, 
\label{eq:velocity}
\end{equation}
where $H^{(W)}_{\alpha}(\bm{k})$ is the $\bm{k}$-derivative of $H^{(W)}(\bm{k})$ in the $\alpha$-direction, evaluated analytically using 
\begin{equation}
H_{\alpha}^{(W)}(\bm{k}) = \partial_{\alpha} H^{(W)}(\bm{k}) =
\sum_{\bm{R}_{e}} e^{i \bm{k} \cdot \bm{R}_{e}} H(\bm{R}_{e}) \cdot (i R^{\alpha}_{e})\,.
\label{eq:wannier_4}
\end{equation}
An appropriate extension of Eq.~(\ref{eq:velocity}) is used for degenerate states~\cite{Yates2007}. 
\subsubsection{Interpolation of the phonon dispersion}
The lattice dynamical properties are first obtained using DFPT on a regular coarse $\bm{q}_{c}$-point grid. Starting from the dynamical matrices $D(\bm{q}_{c})$, we compute the inter-atomic force constants (IFCs) $D(\bm{R}_{p})$ (here, without the mass factor) through a Fourier transform~\cite{Born1954,Gonze1997}, 
\begin{equation}
D(\bm{R}_{p})  =  \frac{1}{N_{p}} \sum_{\bm{q}_{c}} e^{-i\bm{q}_{c} \cdot \bm{R}_{p}} D(\bm{q}_{c}).
\label{eq:IFC1}
\end{equation}
where $N_{p}$ is the number of $\bm{q}_{c}$-points in the coarse grid. If the IFCs are short-ranged, a small set of $D(\bm{R}_{p})$, obtained from dynamical matrices on a coarse $\bm{q}_{c}$-point grid, is sufficient to obtain the dynamical matrix at any desired $\bm{q}$-point with an inverse Fourier transform, 
\begin{equation}
D(\bm{q})  =  \sum_{\bm{R}_{p}} e^{i\bm{q} \cdot \bm{R}_{p}} D(\bm{R}_{p}).
\label{eq:IFC2}
\end{equation}
We obtain the phonon frequencies $\omega_{\nu\bm{q}}$ and displacement eigenvectors $\bm{e}_{\nu\bm{q}}$ by diagonalizing $D(\bm{q})$. 
\subsubsection{Wannier interpolation of the $e$-ph matrix elements}\label{sec:wan_eph}
The key quantities for $e$-ph scattering are the $e$-ph matrix elements $g_{mn\nu}\left(\bm{k},\bm{q}\right)$, which appear above in Eq.~(\ref{eq:e_ph_scattering}). They are given by
\begin{equation}
g_{mn\nu}\left(\bm{k},\bm{q}\right)=\sqrt{\frac{\hbar}{2\omega_{\nu\bm{q}}}}\sum_{\kappa\alpha}\frac{\bm{e}_{\nu\bm{q}}^{\kappa\alpha}}{\sqrt{M_{\kappa}}}\left\langle \psi_{m{\bm{k}+\bm{q}}}\left|\partial_{\bm{q},\kappa\alpha}V\right| \psi_{n\bm{k}}\right\rangle,
\label{eq:eph_mat}
\end{equation}
where $|\psi_{n\bm{k}}\rangle$ and $|\psi_{m\bm{k}+\bm{q}}\rangle$ are the wavefunctions of the initial and final Bloch states, respectively, and $\partial_{\bm{q},\kappa\alpha}V$ is the perturbation potential due to lattice vibrations, computed as the variation of the Kohn-Sham potential $V$ with respect to the atomic displacement of atom $\kappa$ (with mass $M_{\kappa}$) in the Cartesian direction $\alpha$:
\begin{equation}
\partial_{\bm{q}, \kappa \alpha} V=\sum_{\bm{R}_{p}} e^{i \bm{q} \cdot \bm{R}_{p}} \frac{\partial V}{\partial R_{p \kappa \alpha}}.
\label{eq:dvscf}
\end{equation}
We obtain this perturbation potential as a byproduct of the DFPT lattice dynamical calculations at a negligible additional cost. \\
%We obtain this perturbation potential as a byproduct of the DFPT lattice dynamical calculation, which prints it out at negligible additional cost.
%
\indent
We compute the bra-ket in Eq.~(\ref{eq:eph_mat}) directly, using the DFT Bloch states on a coarse $\bm{k}_{c}$-point grid and the perturbation potentials on a coarse $\bm{q}_{c}$-point grid, 
\begin{equation}
\tilde{g}_{mn}^{\kappa\alpha}\left(\bm{k}_{c},\bm{q}_{c}\right)=\left\langle \psi_{m{\bm{k}_{c}+\bm{q}_{c}}}\left|\partial_{\bm{q}_{c},\kappa\alpha}V\right| \psi_{n\bm{k}_{c}}\right\rangle,
\label{eq:eph_mat_bare}
\end{equation}
from which we obtain the $e$-ph matrix elements in the Wannier basis~\cite{Giustino2007,Calandra2010}, $\tilde{g}_{ij}^{\kappa\alpha}\left(\bm{R}_{e},\bm{R}_{p}\right)$, by combining Eq.~(\ref{eq:wannier}) and the inverse transformation of Eq.~(\ref{eq:dvscf}):
%
%To obtained  on fine grid with low cost, we compute the bra-ket in  on ,
%and then obtained the corresponding  using  and ,
\begin{equation}
\begin{aligned}
\tilde{g}_{ij}^{\kappa\alpha}\left(\bm{R}_{e},\bm{R}_{p}\right)   = & \left\langle i \bm{0} \left|\frac{\partial V}{\partial R_{p,\kappa \alpha}} \right|j \bm{R}_{e}\right\rangle \\
 = &\frac{1}{N_{e}N_{p}} \sum_{\bm{k}_{c},\bm{q}_{c}} e^{-i (\bm{k}_{c}\cdot\bm{R}_{e} + \bm{q}_{c}\cdot\bm{R}_{p} )} \tilde{\bm{g}}^{\kappa\alpha, (W)}_{ij}\left(\bm{k}_{c},\bm{q}_{c}\right),
\end{aligned}
\label{eq:eph_mat_wann}
\end{equation}
where  
%\begin{equation}
%\end{equation}
\[
\tilde{\bm{g}}^{\kappa\alpha, (W)}\left(\bm{k}_{c},\bm{q}_{c}\right) = 
\mathcal{U}^{\dagger}(\bm{k}_{c}+\bm{q}_{c}) \tilde{\bm{g}}^{\kappa\alpha}\left(\bm{k}_{c},\bm{q}_{c}\right) \mathcal{U}(\bm{k}_{c})
\]
are the matrix elements in the Wannier gauge.  Similar to the electron Hamiltonian in the Wannier basis, $\tilde{g}_{ij}^{\kappa\alpha}\left(\bm{R}_{e},\bm{R}_{p}\right)$ can be seen as a hopping integral between two localized Wannier functions, one at the origin and one at $\bm{R}_{e}$, due to a perturbation caused by an atomic displacement at $\bm{R}_{p}$. If the interactions are short-ranged in real space, $\tilde{g}$ 
decays rapidly with $|\bm{R}_{e}|$ and $|\bm{R}_{p}|$, and computing it on a small set of $\left(\bm{R}_{e},\bm{R}_{p}\right)$ lattice vectors is sufficient to fully describe the coupling between electrons and lattice vibrations. \\
\indent
The $e$-ph matrix elements at any desired pair of $\bm{k}$- and $\bm{q}$-points can be computed efficiently using the inverse transformation in Eq.~(\ref{eq:eph_mat_wann}), 
\begin{equation}
\begin{aligned}
\tilde{g}_{mn}^{\kappa\alpha}\left(\bm{k},\bm{q}\right)  = & 
\sum_{i,j}U^\dagger_{mi}(\bm{k}+\bm{q})U_{jn}(\bm{k}) \\
& \times \sum_{\bm{R}_{e},\bm{R}_{p}}e^{i (\bm{k}\cdot\bm{R}_{e} + \bm{q}\cdot\bm{R}_{p} )} \tilde{g}_{ij}^{\kappa\alpha}\left(\bm{R}_{e},\bm{R}_{p}\right), 
\end{aligned}
\label{eq:eph_mat_wann2}
\end{equation}
where $U(\bm{k})$ is the matrix used to interpolate the Bloch states in Eq.~(\ref{eq:bloch_state}). \\
\indent
The main requirement of this interpolation approach is that the $e$-ph interactions are short-ranged and the $e$-ph matrix elements in the local basis decay rapidly. Therefore, the $e$-ph interpolation works equally well with localized orbitals other than Wannier functions, as we have shown recently using atomic orbitals~\cite{Agapito2018}. The atomic orbital $e$-ph interpolation is still being extended and tested in the {\sc Perturbo} code, so the current release includes only Wannier interpolation routines. 
%AOs are implemented but will be part of a future release. 
%
%
\subsection{Polar corrections for phonons and $e$-ph interactions}
The assumption that the IFCs and $e$-ph interactions are short-ranged does not hold for polar semiconductors and insulators. In polar materials, the displacement of ions with a non-zero Born effective charge creates dynamical dipoles, and the long-wavelength longitudinal optical (LO) phonon mode induces a macroscopic electric field~\cite{Born1954}. 
The dipole-dipole interactions introduce long-range contributions to the IFCs and dynamical matrices~\cite{Gonze1994},  resulting in the well-known LO-TO splitting in the phonon dispersion at $\bm{q} \rightarrow 0$. 
For this reason, the dynamical matrix interpolation scheme in Eqs.~(\ref{eq:IFC1})-(\ref{eq:IFC2}) cannot provide correct phonon dispersions at small $\bm{q}$ for polar materials. 
To address this issue, a polar correction is typically used~\cite{Gonze1997}, in which the dynamical matrix is separated into two contributions, a short-range part that can be interpolated using the Fourier transformation in Eqs.~(\ref{eq:IFC1})-(\ref{eq:IFC2}), and a long-range part evaluated directly using an analytical formula involving the Born effective charges and the dielectric tensor~\cite{Gonze1997}. {\sc Perturbo} implements this standard approach to include the LO-TO splitting in polar materials. \\
%
%parts: the long-range part due to the dipole-dipole interactions and the remainder, which is short-ranged. The 
%
%short-range part of the dynamical matrix is interpolated using the Fourier transformation in Eqs.~(\ref{eq:IFC1}) and (\ref{eq:IFC2}), while the long-range part is evaluated directly using an analytical formula involving the Born effective charges and the dielectric tensor~\cite{Gonze1997}. 
%
\indent
The field due to the dynamical dipoles similarly introduces long-range $e$-ph contributions $-$ in particular, the Fr{\"o}hlich interaction~\cite{Frohlich1954}, a long-range coupling between electrons and LO phonons. The strength of the Fr{\"o}hlich $e$-ph interaction diverges as $1/q$ for $\bm{q}\rightarrow 0$ in bulk materials. 
As a result, the Wannier interpolation is impractical and usually fails to correctly reproduce the DFPT $e$-ph matrix elements at small $\bm{q}$. Using a scheme analogous to the polar correction for phonon dispersion, one can split the $e$-ph matrix elements into a long-range part due to the dipole field and a short-range part~\cite{Sjakste2015,Verdi2015}.  \\
\indent
We compute the long-range $e$-ph matrix elements by replacing the perturbation potential in Eq.~(\ref{eq:eph_mat_bare}) with the potential of the dipole field, 
\begin{equation}
\begin{aligned}
\tilde{g}_{mn}^{\kappa\alpha, L}\left(\bm{k},\bm{q}\right) = 
& \frac{ie^2}{\epsilon_0\Omega} \sum_{\bm{G} \ne -\bm{q}}{ \frac{ \left[(\bm{q}+\bm{G}) \cdot \bm{Z}_\kappa^{*}\right]_{\alpha}e^{-i\bm{\tau}_\kappa \cdot (\bm{q}+\bm{G})}}{ (\bm{q}+\bm{G})\cdot \bm{\epsilon} \cdot(\bm{q}+\bm{G})} } \\ &
 \times \left\langle {\psi_{m\bm{k}+\bm{q}}} \left| e^{i(\bm{q}+\bm{G})\cdot \bm{r}}  \right| \psi_{n\bm{k}}\right\rangle,
\end{aligned}
\label{eq:e_ph_long}
\end{equation}
where $\bm{Z}^*_\kappa$ and $\bm{\tau}_\kappa$ are the Born effective charge and position of atom $\kappa$ in the unit cell, respectively, while $\Omega$ is the unit cell volume and $\bm{\epsilon}$ the dielectric tensor. 
In practice, the summation over $\bm{G}$ is performed using the Ewald method, by introducing a decay factor $e^{-(\bm{q}+\bm{G})\cdot \bm{\epsilon} \cdot(\bm{q}+\bm{G})/4\varLambda}$ with convergence parameter $\varLambda$, and multiplying each term in the summation by this factor.
It is convenient to evaluate the bra-ket in Eq.~(\ref{eq:e_ph_long}) in the Wannier gauge, in which one can apply the smooth phase approximation $\langle u_{m\bm{k}+\bm{q}}| u_{n\bm{k}}\rangle^{(W)}= \delta_{mn}$, where $u_{n\bm{k}}$ is the periodic part of the Bloch function. Combining Eq.~(\ref{eq:wannier_u}) and the smooth phase approximation in the Wannier gauge, we obtain
\begin{equation}
\left\langle \psi_{m\bm{k}+\bm{q}} \left| e^{i(\bm{q}+\bm{G})\cdot \bm{r}}  \right| \psi_{n\bm{k}}\right\rangle 
= [\mathcal{U}(\bm{k}+\bm{q}) \mathcal{U}^{\dagger}(\bm{k})]_{mn}. 
\label{eq:smooth_phase}
\end{equation}
Using the analytical formula in Eq.~(\ref{eq:e_ph_long}), with the bra-ket computed using Eq.~(\ref{eq:smooth_phase}), the long-range part of the $e$-ph matrix elements can be evaluated directly for any desired values of $\bm{k}$ and $\bm{q}$. Only the short-range part is computed using Wannier interpolation, and the full $e$-ph matrix elements are then obtained by adding together the short- and long-range contributions. \\
\indent
To extend the phonon and $e$-ph polar correction schemes to 2D materials, only small changes to the long-range parts are needed, as discussed in detail in Refs.~\cite{Sohier2016,Sohier2017,Sohier2017a}. 
In particular, the 2D extension of the long-range $e$-ph matrix elements is obtained by replacing in Eq.~(\ref{eq:e_ph_long}) the dielectric tensor $\bm{\epsilon}$ with the effective screening $\bm{\epsilon}_{\text{eff}}(|\bm{q}|)$ of the 2D system and the unit cell volume $\Omega$ with $2A$, where $A$ is the area of the 2D unit cell. Polar corrections for the phonon dispersion and $e$-ph matrix elements are supported in {\sc Perturbo} for both bulk and 2D materials.
%
%
%The polar correction scheme also applies to 2D materials, only small modification to the formula of long-range part is required, which has been discussed in Ref.~\cite{Sohier2016} and~\cite{Sohier2017}. For examples, one can obtain the 2D extension of long-range $e$-ph matrix elements by replacing in Eq.~(\ref{eq:e_ph_long}) the dielectric tensor $\bm{\epsilon}$ with the effective screening $\bm{\epsilon}_{\text{eff}}(|\bm{q}|)$ of the 2D system and replacing the unit cell volume $\Omega$ with $2A$, where $A$ is the area of the 2D unit cell. The polar corrections of phonon dispersion and $e$-ph matrix elements for both 2D and 3D materials are supported in {\sc Perturbo}. 
%
%
%workflow figures
\begin{figure*}[!htb]
\centering
\includegraphics[width=2\columnwidth]{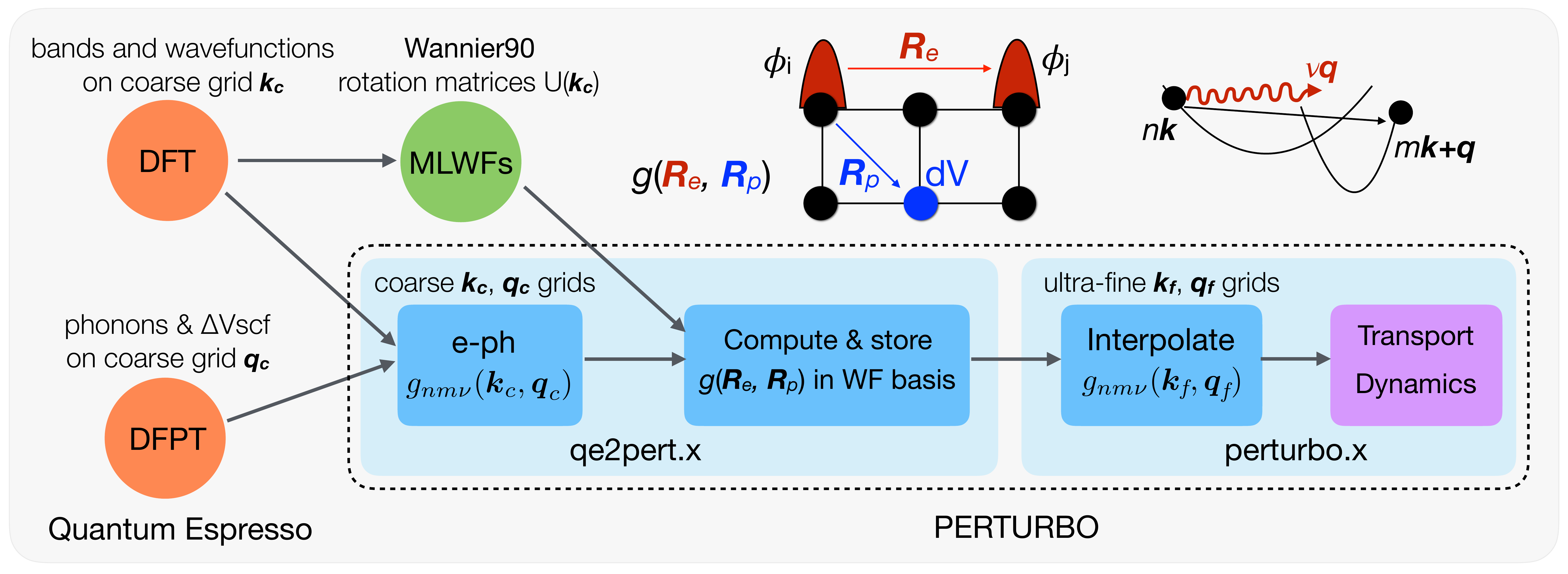}
\caption{
Workflow of the {\sc Perturbo} code. Before running {\sc Perturbo}, the user performs DFT calculations on the system of interest to obtain the Bloch states on a coarse $\bm{k}_{c}$-point grid and DFPT calculations to obtain the lattice dynamical properties and perturbation potentials on a coarse $\bm{q}_{c}$-point grid. The \texttt{qe2pert.x} executable of {\sc Perturbo} computes the $e$-ph matrix elements on coarse $\bm{k}_{c}$- and $\bm{q}_{c}$-point grids, from which the $e$-ph matrix elements in the localized Wannier basis are obtained with the rotation matrices from {\sc Wannier90}. The core executable \texttt{perturbo.x} is then employed to interpolate the band structure, phonon dispersion, and $e$-ph matrix elements on ultra-fine $\bm{k}_{f}$- and $\bm{q}_{f}$-point grids, and to perform charge transport and carrier dynamics calculations.}
\label{fig:workflow}
\end{figure*}
\section{Capabilities and workflow} \label{sec:cap}
\subsection{Code organization and capabilities}
{\sc Perturbo} contains two executables: a core program \texttt{perturbo.x} and the program \texttt{qe2pert.x}, which interfaces the QE code and \texttt{perturbo.x}, as shown in Fig.~\ref{fig:workflow}. The current release supports DFT and DFPT calculations with norm-conserving or ultrasoft pseudopotentials, with or without SOC; it also supports the Coulomb truncation for 2D materials~\cite{Sohier2017a}.
%scalar and fully relativistic pseudopotentials (with SOC), and also the
%and Coulomb truncation for 2D materials~\cite{Sohier2017a}. 
The current features include calculations of: 
\begin{enumerate}[1.]
\item Band structure, phonon dispersion, and $e$-ph matrix elements on arbitrary BZ grids or paths.
\item The $e$-ph scattering rates, relaxation times, and electron mean free paths for electronic states with any band and $\bm{k}$-point.  
\item Electrical conductivity, carrier mobility, and Seebeck coefficient using the RTA or the iterative solution of the BTE.
\item Nonequilibrium dynamics, such as simulating the cooling and equilibration of excited carriers via interactions with phonons.
\end{enumerate}%
Several features of {\sc Perturbo}, including the nonequilibrium dynamics, are unique and not available in other existing codes. 
%The code can carry out many workflows (e.g., ultrafast carrier dynamics and transport in 2D materials, among others) that are not possible with other existing codes.
Many additional features, currently being developed or tested, will be added to the list in future releases. \\
\indent
{\sc Perturbo} stores most of the data and results in HDF5 file format, including all the main results of \texttt{qe2pert.x}, the TDF and other files generated in the transport calculations, and the nonequilibrium electron distribution computed in \texttt{perturbo.x}. The use of the HDF5 file format improves the portability of  the results between different computing systems and is also convenient for post-processing using high-level languages, such as Python and Julia. \\
\indent
Installation and usage of {\sc Perturbo} and an up-to-date list of supported features are documented in the user manual distributed along with the source code package, and can also be found on the code website~\cite{perturbo}. \\
\subsection{Computational workflow}
Figure~\ref{fig:workflow} summarizes the workflow of {\sc Perturbo}. Before running {\sc Perturbo}, the user needs to carry out DFT and DFPT calculations with the QE code, and Wannier function calculations with {\sc Wannier90}. In our workflow, we first carry out DFT calculations to obtain the band energies and Bloch wavefunctions on a coarse grid with points $\bm{k}_{c}$. 
A regular Monkhorst-Pack (MP) $\bm{k}_{c}$-point grid centered at $\Gamma$ and Bloch states for all $\bm{k}_{c}$-points in the first BZ are required.
We then construct a set of Wannier functions from the Bloch wavefunctions using the {\sc Wannier90} code. Only the rotation matrices $\mathcal{U}$ that transform the DFT Bloch wavefunctions to the Wannier gauge and the center of the Wannier functions are required as input to {\sc Perturbo}. 
We also perform DFPT calculations to obtain the dynamical matrices and $e$-ph perturbation potentials on a coarse MP grid with points $\bm{q}_{c}$. 
In the current version, the electron $\bm{k}_{c}$-point grid and phonon $\bm{q}_{c}$-point grid need to be commensurate.
Since DFPT is computationally demanding, we carry out the DFPT calculations only for $\bm{q}_{c}$-points in the irreducible wedge of the BZ, and then obtain the dynamical matrices and perturbation potentials in the full BZ grid using space group and time reversal symmetries (see Sec.~\ref{sec:coarse_g}). \\
\indent
The executable \texttt{qe2pert.x} reads the results from DFT and DFPT calculations, including the Bloch states $| \psi_{n\bm{k}_{c}} \rangle$, dynamical matrices $D(\bm{q}_{c})$ and $e$-ph perturbation potentials $\partial_{\bm{q}_{c}, \kappa \alpha} V$, and then computes the electron Hamiltonian in the Wannier basis [Eq.~(\ref{eq:wannier_2})] and the IFCs [Eq.~(\ref{eq:IFC1})]. 
It also computes the $e$-ph matrix elements on coarse grids and transforms them to the localized Wannier basis using Eqs.~(\ref{eq:eph_mat_bare})-(\ref{eq:eph_mat_wann}). To ensure that the same Wannier functions are used for the electron Hamiltonian and $e$-ph matrix elements, we use the same DFT Bloch states $| \psi_{n\bm{k}_{c}} \rangle$ and $\mathcal{U}(\bm{k}_{c})$ matrices for the calculations in Eqs.~(\ref{eq:wannier_2}) and~(\ref{eq:eph_mat_bare})-(\ref{eq:eph_mat_wann}). Following these preliminary steps, \texttt{qe2pert.x} outputs all the relevant data to an HDF5 file, which is the main input for \texttt{perturbo.x}. \\
\indent
The executable \texttt{perturbo.x} reads the data computed by \texttt{qe2pert.x} and carries out the transport and dynamics calculations discussed above. To accomplish these tasks, \texttt{perturbo.x} interpolates the band structure, phonon dispersion, and $e$-ph matrix elements on fine $\bm{k}$- and $\bm{q}$-point BZ grids, and uses these quantities in the various calculations.
%
%
%The \texttt{perturbo} package contains two executables.
%support  SOC,  norm-conserving pseudo,  ultrasoft (not extensively test, use with cautions)
%[Only input, U matrix, center of the wannier function.] \cite{Pizzi2019}
%* interpolation of electron, phonon and electron-phonon coupling matrix elements
%* electron scattering rate and mean free path
%*  using RTA and iterative solution
%* hot carrier cooling . 
%used in hot carrier cooling and high field transport,  The current release only support hot %carrier cooling. 
%%\subsection{
%Data flow and %HDF5 file format
%we will develop post-processing using high level langugage: python or julia. 
%[capabilities of our code. ]
%[how to compile and run]
%
%
\section{Technical aspects} \label{sec:tech}
\subsection{$e$-ph matrix elements on coarse grids} \label{sec:coarse_g}
As discussed in Sec.~\ref{sec:wan_eph}, we compute directly the $e$-ph matrix elements in Eq.~(\ref{eq:eph_mat_bare}) using the DFT states on a coarse $\bm{k}_{c}$-point grid and the perturbation potentials on a coarse $\bm{q}_{c}$-point grid. It is convenient to rewrite Eq.~(\ref{eq:eph_mat_bare}) in terms of lattice-periodic quantities,
%The most time-consuming part of \texttt{qe2pert.x} is the calculations of $g_{mn}^{\kappa\alpha}\left(\bm{k}_{c},\bm{q}_{c}\right)$ in , which can be rewritten  
\begin{equation}
\tilde{g}_{mn}^{\kappa\alpha}\left(\bm{k}_{c},\bm{q}_{c}\right)=\left\langle u_{m{\bm{k}_{c}+\bm{q}_{c}}}\left|\partial_{\bm{q}_{c},\kappa\alpha}v\right| u_{n\bm{k}_{c}}\right\rangle,
\label{eq:eph_mat_bare2}
\end{equation}
where $| u_{n\bm{k}_{c}}\rangle$ is the lattice periodic part of the Bloch wavefunction and  
$\partial_{\bm{q}_{c},\kappa\alpha}v = e^{-i\bm{q}_{c} \cdot \bm{r}} \partial_{\bm{q}_{c},\kappa\alpha}V$
is the lattice-periodic perturbation potential. 
%and  are the lattice periodic part of the Bloch wavefunction $| \psi_{n\bm{k}_{c}} \rangle$ and $e$-ph perturbation potential $\partial_{\bm{q}_{c},\kappa\alpha}V$, respectively, e.g. 
%
%
Since we compute only $| u_{n\bm{k}_{c}}\rangle$ on the coarse grid of the first BZ, $| u_{m\bm{k}_{c}+\bm{q}_{c}}\rangle$ may not be available because $\bm{k}_{c}+\bm{q}_{c}$ may be outside of the first BZ. 
However, by requiring the $\bm{q}_{c}$-point grid to be commensurate with and smaller than (or equal to) the $\bm{k}_{c}$-point grid, we can satisfy the relationship $\bm{k}_{c}+\bm{q}_{c} = \bm{k}^{\prime}_{c} + \bm{G}_{0}$, where $\bm{k}^{\prime}_{c}$ is on the coarse grid and $\bm{G}_{0}$ is a reciprocal lattice vector. 
Starting from $|u_{m\bm{k}^{\prime}_{c}}\rangle = \sum_{\bm{G}} c_{\bm{k}^{\prime}_{c}}(\bm{G}) e^{i\bm{G} \cdot \bm{r}}$, we can thus obtain $| u_{m\bm{k}_{c}+\bm{q}_{c}}\rangle$ with negligible computational cost as
\begin{equation}
|u_{m\bm{k}_{c}+\bm{q}_{c}}\rangle = 
e^{-i\bm{G}_{0} \cdot \bm{r}}| u_{m\bm{k}^{\prime}_{c}}\rangle
=  \sum_{\bm{G}} c_{\bm{k}^{\prime}_{c}}(\bm{G}) e^{i(\bm{G}-\bm{G}_{0}) \cdot \bm{r}}.
\end{equation}
%
%We require the $\bm{q}_{c}$-grid to be commensurable to and smaller than the $\bm{k}_{c}$ grid, thus we can assure $\bm{k}_{c}+\bm{q}_{c} = \bm{k}^{\prime}_{c} + \bm{G_{0}}$ where $\bm{k}^{\prime}_{c}$ is on the coarse grid and $\bm{G_{0}}$ is a reciprocal lattice vector. 
%
%Assuming $\bm{k}_{c}+\bm{q}_{c} = \bm{k}^{\prime}_{c} + \bm{G_{0}}$ where $\bm{k}^{\prime}_{c}$ is within the first BZ and  $\bm{G_{0}}$ is a reciprocal lattice vector, and 
%$|\bm{u}_{m,\bm{k}^{\prime}_{c}}\rangle = \sum_{G} c_{\bm{k}^{\prime}_{c}}(\bm{G}) e^{i\bm{G}\bm{r}}$, then we can obtain $| \bm{u}_{m,\bm{k}_{c}+\bm{q}_{c}}\rangle$ as 
%\begin{equation}
%|\bm{u}_{m,\bm{k}_{c}+\bm{q}_{c}}\rangle = 
%e^{-\bm{G_{0}}\bm{r}}| \bm{u}_{m,\bm{k}^{\prime}_{c}}\rangle
%=  \sum_{G} c_{\bm{k}^{\prime}_{c}}(\bm{G}) e^{i(\bm{G}-\bm{G_{0}})\bm{r}}.
%\end{equation}
%which requires negligible cost. 
%
\indent
The lattice-periodic perturbation potential $\partial_{\bm{q}_{c},\kappa\alpha}v$ consists of multiple terms, and can be  divided into a local and a non-local part~\cite{DalCorso2001}. 
The non-local part includes the perturbation potentials due to the non-local terms of pseudopotentials, which typically includes the Kleinman-Bylander projectors and SOC terms. 
The local part includes the perturbations to the local part of the pseudopotentials as well as the self-consistent potential contribution.
The latter, denoted as $\partial_{\bm{q}_{c},\kappa\alpha}v_{\rm sc}(\bm{r})$, accounts for the change in the Hartree and exchange-correlation potentials in response to the atomic displacements.
While the pseudopotential contributions, both local and non-local, can be evaluated efficiently for all $\bm{q}_{c}$-points with the analytical formula given in Ref.~\cite{DalCorso2001}, 
the self-consistent contribution $\partial_{\bm{q}_{c},\kappa\alpha}v_{\rm sc}(\bm{r})$ is computed and stored in real space using expensive DFPT calculations. This step is the main bottleneck of the entire $e$-ph computational workflow. \\
%
%
%The use of pseudopotentials introduces non-local part, namely the change in the non-local part of pseudopotentials which includes the Kleinman-Bylander terms and spin-orbital coupling terms. 
%The local part of $\partial_{\bm{q}_{c},\kappa\alpha}v$ contains the change in the local part of pseudopotentials and a self-consistent contribution
%
%, $\partial_{\bm{q}_{c},\kappa\alpha}v_{sc}$, that account for the change in the Hartree and exchange-correlation potential. 
%
%The contribution to perturbation potential from the change in the local and non-local part of pseudopotentials can be computed efficiently with the analytical formula given in Ref.~\cite{DalCorso2001} for arbitrary $\bm{q}-$point. The local self-consistent contribution $\partial_{\bm{q}_{c},\kappa\alpha}v_{sc}$ need to be evaluated in real space using DFPT calculations. \\
%
%Because the Kohn-Sham potential $V$ has non-local parts due to the use of pseudopotentials in DFT implementation,  contains . The local part of $\partial_{\bm{q}_{c},\kappa\alpha}v$ also \\
%
%\indent
%Note that Eq.~(\ref{eq:rot_dvscf}) only works if $\partial_{\bm{q}_{c},\kappa\alpha}v$ is local and scalar function. In case of non-magnetic and SOC, then we need to apply rotation in spin space as well. for non-local port of the dvscf, computed directly without symmetry operation. \\
%
%DFPT calculations are computationally expensive. 
%
\indent
To improve the efficiency, we compute the self-consistent contribution with DFPT only for $\bm{q}_{c}$-points in the irreducible BZ wedge, and then unfold it to the equivalent points $\mathcal{S}\bm{q}_{c}$ in the full BZ using symmetry operations~\cite{MARADUDIN1968,Agapito2018}. 
For non-magnetic systems,  $\partial_{\bm{q}_{c},\kappa\alpha}v_{\rm sc}(\bm{r})$ is a scalar function, so we can obtain the self-consistent contribution at $\mathcal{S}\bm{q}_{c}$ by rotating $\partial_{\bm{q}_{c},\kappa\alpha}v_{\rm sc}(\bm{r})$~\cite{Agapito2018}:
%For scalar function $\partial_{\bm{q}_{c},\kappa\alpha}v_{\rm sc}(\bm{r})$,  
%
\begin{equation}
\begin{aligned}
\partial_{\mathcal{S}\bm{q}_{c}, \kappa \alpha} v_{\rm sc}(\bm{r})  = &  \sum_{\kappa^{\prime} \beta} e^{i \bm{q}_{c} \cdot \bm{\tau}_{\kappa^{\prime}}-i \mathcal{S}\bm{q}_{c} \cdot \bm{\tau}_{\kappa}} \\ & \times \left[\mathcal{S}^{-1}\right]_{\beta \alpha} \partial_{\bm{q}_{c}, \kappa^{\prime} \beta} v_{\rm sc}\left(\{\mathcal{S} | \bm{t}\}^{-1} \bm{r}\right),
\end{aligned}
\label{eq:rot_dvscf}
\end{equation}
where $\{\mathcal{S}|\bm{t} \}$ is a space group symmetry operation of the crystal. A detailed derivation of Eq.~(\ref{eq:rot_dvscf}) can be found in Appendix C of Ref.~\cite{Agapito2018}. In addition to space group operations, time reversal symmetry is also used for non-magnetic systems via
\begin{equation}
\partial_{-\bm{q}_{c}, \kappa \alpha} v_{\rm sc}(\bm{r}) = [\partial_{\bm{q}_{c}, \kappa \alpha} v_{\rm sc}(\bm{r})]^{*}, 
\end{equation}
since the time reversal symmetry operator for a scalar function is the complex conjugation operator. 
We emphasize that Eq.~(\ref{eq:rot_dvscf}) is only used to unfold the self-consistent contribution of the perturbation potential, while all the terms due to the pseudopotentials are computed directly, without using symmetry, for all the $\bm{q}_{c}$-points in the coarse grid. \\
%
%symmetry and time reversal symmetry 
%The use of space group symmetry and time reversal symmetry greatly reduced the computational cost for the calculations of $e$-ph matrix elements on coarse grids. \\
%
%
\begin{figure}[!htb]
\centering
\includegraphics[width=0.92\columnwidth]{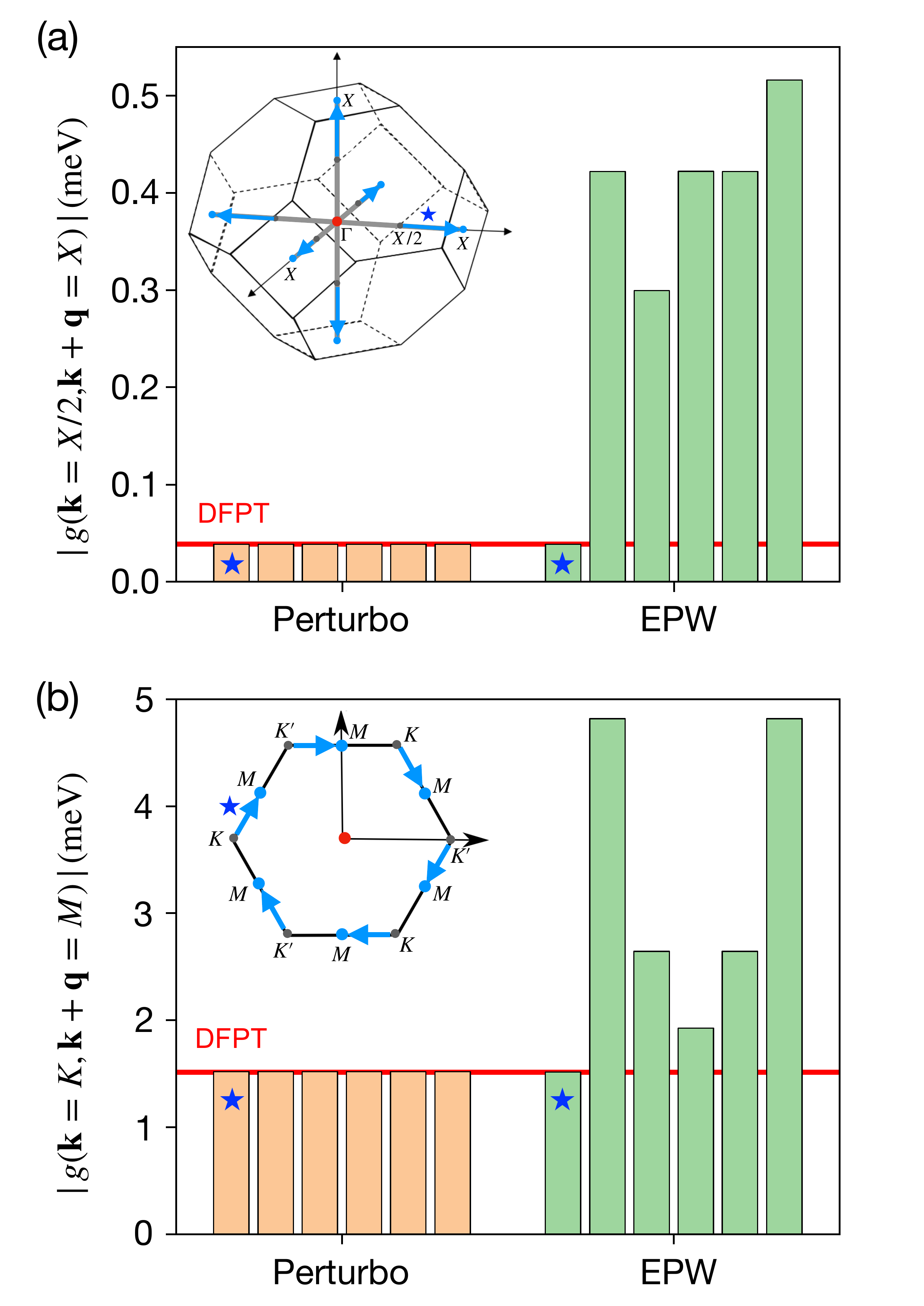}
\caption{Absolute value of the gauge-invariant $e$-ph matrix elements, $\left|g\left(\bm{k},\bm{q}\right)\right|$ in Eq.~(\ref{eq:g_abs}), computed with {\sc Perturbo} (orange) and {\sc EPW} (green) in (a) silicon and (b) monolayer MoS$_{2}$. SOC is included in both cases. The six bars are $\left|g\left(\bm{k},\bm{q}\right)\right|$ values for six equivalent $(\bm{k}, \bm{q})$ pairs connected by symmetry, which are shown as blue arrows in the inset. The result with $\bm{q}$ in the irreducible wedge (labeled with a blue star) is computed directly with the DFPT perturbation potential, while the others are obtained by applying symmetry operations either on the perturbation potentials ({\sc Perturbo}) or on the wavefunctions ({\sc EPW}). The red horizontal line shows the benchmark values computed directly from DFPT for all $(\bm{k}, \bm{q})$ pairs.}
\label{fig:eph-cg}
\end{figure}
%
%\indent
%It is worth mention that an alternative approach to obtain $e$-ph matrix elements for reducible $\bm{q}_{c}$  using symmetry  is to rotate the wavefunctions rather than 
%
%there is an alternative approach to unfold the 
%
%
\indent
An alternative approach to obtain the $e$-ph matrix elements in the full BZ starting from results in the irreducible wedge is to rotate the wavefunctions instead of the perturbation potential: 
\begin{equation}
\begin{aligned}
\langle u_{m{\bm{k}_{c}+\mathcal{S}\bm{q}_{c}}} & |\partial_{\mathcal{S}\bm{q}_{c},\kappa\alpha}v |  u_{n\bm{k}_{c}} \rangle
 =  \\ 
& \langle u_{m{\bm{k}_{c}+\mathcal{S}\bm{q}_{c}}} | \mathcal{D}_{\{\mathcal{S} | \bm{t}\}}^\dagger \partial_{\bm{q}_{c},\kappa\alpha}v \,\mathcal{D}_{\{\mathcal{S} | \bm{t}\}} |  u_{n\bm{k}_{c}} \rangle,
\label{eq:eph_mat_rotwfc}
\end{aligned}
\end{equation}
where $\mathcal{D}_{\{\mathcal{S} | \bm{t}\}}$ is the symmetry operator acting on the wavefunction. In this approach (not employed in {\sc Perturbo}), the perturbation potentials are needed only for $\bm{q}_c$ in the irreducible wedge~\cite{Giustino2007}. It is important to keep in mind that the wavefunctions are spinors in non-collinear calculations, so the symmetry operators $\mathcal{D}_{\{\mathcal{S} | \bm{t}\}}$ should act both on the spatial coordinate and in spin space. Neglecting the rotation in spin space would lead to significant errors in the computed $e$-ph matrix elements, especially in calculations with SOC.  \\
\indent
To benchmark our implementation, we compare the $e$-ph matrix elements obtained using symmetry operations to those from direct DFPT calculations. The absolute value of the $e$-ph matrix elements, $\left|g\left(\bm{k},\bm{q}\right)\right|$, is computed in gauge-invariant form for each phonon mode (with index $\nu$) by summing over bands:
\begin{equation}
\left|g_{\nu}\left(\bm{k},\bm{q}\right)\right| = \sqrt{\sum_{mn}\left|g_{mn\nu}\left(\bm{k},\bm{q}\right)\right|^{2}/N_{b}},
\label{eq:g_abs}
\end{equation}
where $m$, $n$ are band indices for the selected $N_{b}$ bands. 
We perform the comparison with direct DFPT calculations for silicon and monolayer MoS$_{2}$ as examples of a bulk and a 2D material, respectively. We include SOC in both cases. For silicon, we choose $\bm{k}=X/2$, $\bm{k}+\bm{q} = X$ and compute $\left|g\left(\bm{k},\bm{q}\right)\right|$ in Eq.~(\ref{eq:g_abs}) using the four highest valence bands; for monolayer MoS$_{2}$, we choose $\bm{k} = K$, $\bm{k}+\bm{q} = M$ and compute $\left|g\left(\bm{k},\bm{q}\right)\right|$ for the two lowest conduction bands with 2D Coulomb truncation turned off. For both silicon and monolayer MoS$_{2}$, we compute $\left|g\left(\bm{k},\bm{q}\right)\right|$ for all the six \emph{equivalent} $(\bm{k}, \bm{q})$ pairs connected by space group and time reversal symmetry [see the inset of Fig.~\ref{fig:eph-cg}(a, b)].
As a benchmark, DFPT calculations are carried out to evaluate directly $\left|g\left(\bm{k},\bm{q}\right)\right|$ for all the six $(\bm{k}, \bm{q})$ pairs. The results shown in Fig.~\ref{fig:eph-cg} are for the lowest acoustic mode, though the results for the other modes show similar trends.
The $\left|g\left(\bm{k},\bm{q}\right)\right|$ values computed with DFPT are identical for the six equivalent $(\bm{k}, \bm{q})$ pairs (see the red horizontal line in Fig.~\ref{fig:eph-cg}), which is expected based on symmetry.  
In the {\sc Perturbo} calculation, only the $(\bm{k}, \bm{q})$ pair with $\bm{q}$ in the irreducible wedge is computed directly with the perturbation potential from DFPT, 
while results for the other five equivalent $(\bm{k}, \bm{q})$ pairs are obtained by rotating the self-consistent perturbation potential. 
The results obtained with this approach match to a high accuracy the DFPT benchmark results, which validates the perturbation potential rotation approach implemented in {\sc Perturbo}. \\
\indent
For comparison, we show in Fig.~\ref{fig:eph-cg} the results from the alternative approach of rotating the wavefunctions, as implemented in the {\sc EPW} code (version 5.2)~\cite{Ponce2016}. 
Surprisingly, using the {\sc EPW} code only the $\left|g\left(\bm{k},\bm{q}\right)\right|$ value for the $(\bm{k}, \bm{q})$ pair containing the irreducible $\bm{q}$-point agrees with the DFPT benchmark, while all other $\left|g\left(\bm{k},\bm{q}\right)\right|$ values for $\bm{q}$-points obtained using symmetry operations show significant errors. 
We stress again that all the results in Fig.~\ref{fig:eph-cg} are computed with SOC. We have verified that, in the absence of SOC, both {\sc Perturbo} and {\sc EPW} produce results in agreement with DFPT. The likely reason for the failure of {\sc EPW} in this test is that in {\sc EPW} the wavefunctions are rotated as scalars even in the presence of SOC, rather than as spinors as they should (that is, the rotation in spin space is missing). 
% in the presence of SOC.
%of the discrepancy is that  neglects the rotation of wavefunctions in spin space . 
Further investigation of this discrepancy is critical, since the large errors in {\sc EPW} for the coarse-grid $e$-ph matrix elements will propagate to the interpolated matrix elements on fine grids, giving incorrect results in calculations including SOC carried out with {\sc EPW}~\cite{Gaddemane2019}.
\subsection{Wigner-Seitz supercell for Wannier interpolation}
As discussed in Sec.~\ref{sec:wannier}, the DFT Bloch states obtained at $\bm{k}_{c}$-points on a regular BZ grid are used to construct the Wannier functions. A discrete BZ grid implies a Born-von Karman (BvK) boundary condition in real space, so that an $N \times N \times N$ $\bm{k}_{c}$-point grid corresponds (for simple cubic, but extensions are trivial) to a BvK supercell of size $N\!a \times N\!a \times N\!a$, where $a$ is the lattice constant of the unit cell. 
If we regard the crystal as made up by an infinite set of BvK supercells at lattice vectors $\bm{T}_{e}$, we can label any unit cell in the crystal through its position $\bm{T}_{e} + \bm{R}_{e}$, where $\bm{R}_{e}$ is the unit cell position in the BvK supercell. 
%$\bm{T}_{e}$ denotes the lattice vector of the BvK supercell and 
Because of the BvK boundary condition, the Bloch wavefunctions are truly periodic functions over the BvK supercell, and the Wannier function $| n \bm{R}_{e}\rangle$ obtained using Eq.~(\ref{eq:wannier}) is actually the superposition of images of  the Wannier function in all the BvK supercells. Therefore, we can write
%\begin{equation}
$|n \bm{R}_{e}\rangle = \sum_{\bm{T}_{e}} |n, \bm{R}_{e}+\bm{T}_{e}\rangle^{0}$, where $|n, \bm{R}_{e}+\bm{T}_{e}\rangle^{0}$ denotes the image of the Wannier function in the BvK supercell at $\bm{T}_{e}$. 
%\end{equation}
Similarly, the electron Hamiltonian computed using Eq.~(\ref{eq:wannier_2}) can be expressed as
\begin{equation}
H_{nn^{\prime}}(\bm{R}_{e}) = \sum_{\bm{T}_{e}} H^{0}_{nn^{\prime}}(\bm{R}_{e}+\bm{T}_{e}).
\label{eq:wann_ws}
\end{equation}
The hopping matrix elements $H^{0}_{nn^{\prime}}(\bm{R}_{e}+\bm{T}_{e})$ usually decay rapidly as the distance between two image Wannier functions increases.
The BvK supercell should be large enough to guarantee that only the hopping term between the two Wannier function images with the shortest distance is significant, while all other terms in the summation over $\bm{T}_{e}$ in Eq.~(\ref{eq:wann_ws}) are negligible. \\
\indent
We use this ``least-distance'' principle to guide our choice of a set of  $\bm{\tilde{R}}_{e}$ vectors for the Wannier interpolation. For each Hamiltonian matrix element labeled by $(n, n^{\prime}, \bm{R}_{e})$ in Eq.~(\ref{eq:wann_ws}), we compute the distance $d = |\bm{T}_{e}+\bm{R}_{e}+\bm{\tau}_{n^{\prime}} - \bm{\tau}_{n}|$, with $\bm{\tau}_{n}$ the position of the Wannier function center in the unit cell, and find the vector $\bm{T}_{e}^{0}$ giving the minimum distance.
The set of vectors $\bm{\tilde{R}}_{e} = \bm{R}_{e}+\bm{T}_{e}^{0}$ is then selected to construct the Wigner-Seitz supercell used in the Wannier interpolation. 
We compute $H_{nn^{\prime}}(\bm{\tilde{R}}_{e})$ using Eq.~(\ref{eq:wannier_2}) and use it to interpolate the band structure with Eq.~(\ref{eq:wannier_3}). 
Note that the same strategy to construct the Wigner-Seitz supercell is also used in the latest version of {\sc Wannier90}~\cite{Pizzi2020}. \\
\indent
Similarly, we choose a set of least-distance $\bm{\tilde{R}}_{p}$ vectors for the interpolation of the phonon dispersion, and separately determine least-distance $\bm{\tilde{R}}_{e}$ and $\bm{\tilde{R}}_{p}$ pairs for the interpolation of the $e$-ph matrix elements in Eqs.~(\ref{eq:eph_mat_wann})-(\ref{eq:eph_mat_wann2}).
%
%, we choose a set of least-distance $\bm{\tilde{R}}_{e}$ and $\bm{\tilde{R}}_{p})$ vectors that give minimum $|\bm{\tilde{R}}_{e}+\bm{\tau}_{n^{\prime}} - \bm{\tau}_{n}|$ and $|\bm{\tilde{R}}_{p}+\bm{\tau}_{\kappa} - \bm{\tau}_{n}|$, respectively. 
%we choose a set of least-distance $\bm{\tilde{R}}_{p})$ vectors that gives the minimum distance between the position of the displaced atom and 
%$\bm{\tilde{R}}_{e}$ and $\bm{\tilde{R}}_{p})$ vectors 
 %and a set of least-distance $(\bm{\tilde{R}}_{e}$ and$\bm{\tilde{R}}_{p})$ vector pairs for the interpolation of  the $e$-ph matrix elements. 
%[For $g(Re, Rp)$, an more sophistical scheme is possible, describe it, but it provide very little improvement but hurt performance. ] 
%
%
\subsection{Brillouin zone sampling and integration}
Several computational tasks carried out by {\sc Perturbo} require integration in the BZ. Examples include scattering rate and TDF calculations, in Eqs.~(\ref{eq:scatter_rate}) and~(\ref{eq:cond_tdf2}), respectively, and the iterative BTE solution in Eq.~(\ref{eq:BTE_iter}). In {\sc Perturbo}, we adopt different BZ sampling and integration approaches for different kinds of calculations.  For transport calculations, we use the tetrahedron method~\cite{Bloechl1994} for the integration over $\bm{k}$ in Eq.~(\ref{eq:cond_tdf2}). 
We sample $\bm{k}$-points in the first BZ using a regular MP grid centered at $\Gamma$, and divide the BZ into small tetrahedra by connecting neighboring $\bm{k}$-points. 
The integration is first performed inside each tetrahedron, and the results are then added together to compute the BZ integral. 
To speed up these transport calculations, we set up a user-defined window spanning a small energy range (typically $\sim$0.5 eV) near the band edge in semiconductors or Fermi level in metals, and restrict the BZ integration to the $\bm{k}$-points with electronic states in the energy window. 
Since only states within a few times the thermal energy $k_{B}T$ of the band edge in semiconductors (or Fermi energy in metals) contribute to transport, including in the tetrahedron integration only $\bm{k}$-points in the relevant energy window greatly reduces the computational cost. \\
\indent
To compute the $e$-ph scattering rate for states with given bands and $\bm{k}$-points, we use the Monte Carlo integration as the default option. We sample random $\bm{q}$-points in the first BZ to carry out the summation over $\bm{q}$ in Eq.~(\ref{eq:scatter_rate}) and obtain the scattering rate. 
One can either increase the number of random $\bm{q}$-points until the scattering rate is converged or average the results from independent samples. 
Note that the energy broadening parameter used to approximate the $\delta$ function in Eq.~(\ref{eq:BTE_W}) is important for the convergence of the scattering rate, so the convergence with respect to both the $\bm{q}$-point grid and broadening needs to be checked carefully~\cite{Zhou2016}. 
{\sc Perturbo} supports random sampling of the $\bm{q}$-points with either a uniform or a Cauchy distribution; a user-defined $\bm{q}$-point grid can also be specified in an external file and used in the calculation. \\
\indent
In the carrier dynamics simulations and in the iterative BTE solution in Eq.~(\ref{eq:BTE_iter}), the $\bm{k}$- and $\bm{q}$-points should both be on a regular MP grid centered at $\Gamma$. 
The two grids should be commensurate, with the size of the $\bm{q}$-point grid smaller than or equal to the size of the $\bm{k}$-point grid. 
This way, we satisfy the requirement in Eq.~(\ref{eq:BTE_iter}) that each $(\bm{k}$+$\bm{q})$-point is also on the $\bm{k}$-point grid. 
To perform efficiently the summation in Eq.~(\ref{eq:BTE_iter}) for all the $\bm{k}$-points, we organize the scattering probability in Eq.~(\ref{eq:BTE_W}) using ($\bm{k}$, $\bm{q}$) pairs. 
We first determine a set of bands and $\bm{k}$-points for states inside the energy window. We then find all the possible scattering processes in which both the initial and final states are in the selected set of $\bm{k}$-points, and the  phonon wave vector connecting the two states is on the $\bm{q}$-point grid. 
The scattering processes are indexed as ($\bm{k}$, $\bm{q}$) pairs; their corresponding $e$-ph matrix elements are computed and stored, and then retrieved from memory during the iteration process.
%
%\subsubsection{setup k-grid for transport, tetrahedron integration }
%for transport, we use regular $\bm{k}$ grid, and 
%\subsubsection{ (k,q) pair for iterative, dynamics}
%determine $\bm{q}$-grid from selected $\bm{k}$-grid. Note that the scattering rate of  $\bm{k}$ at near the cutoff energy is not accurate since some scattering channel is omitted. 
%
%
\begin{figure}[!hbh]
\centering
\includegraphics[width=0.84\columnwidth]{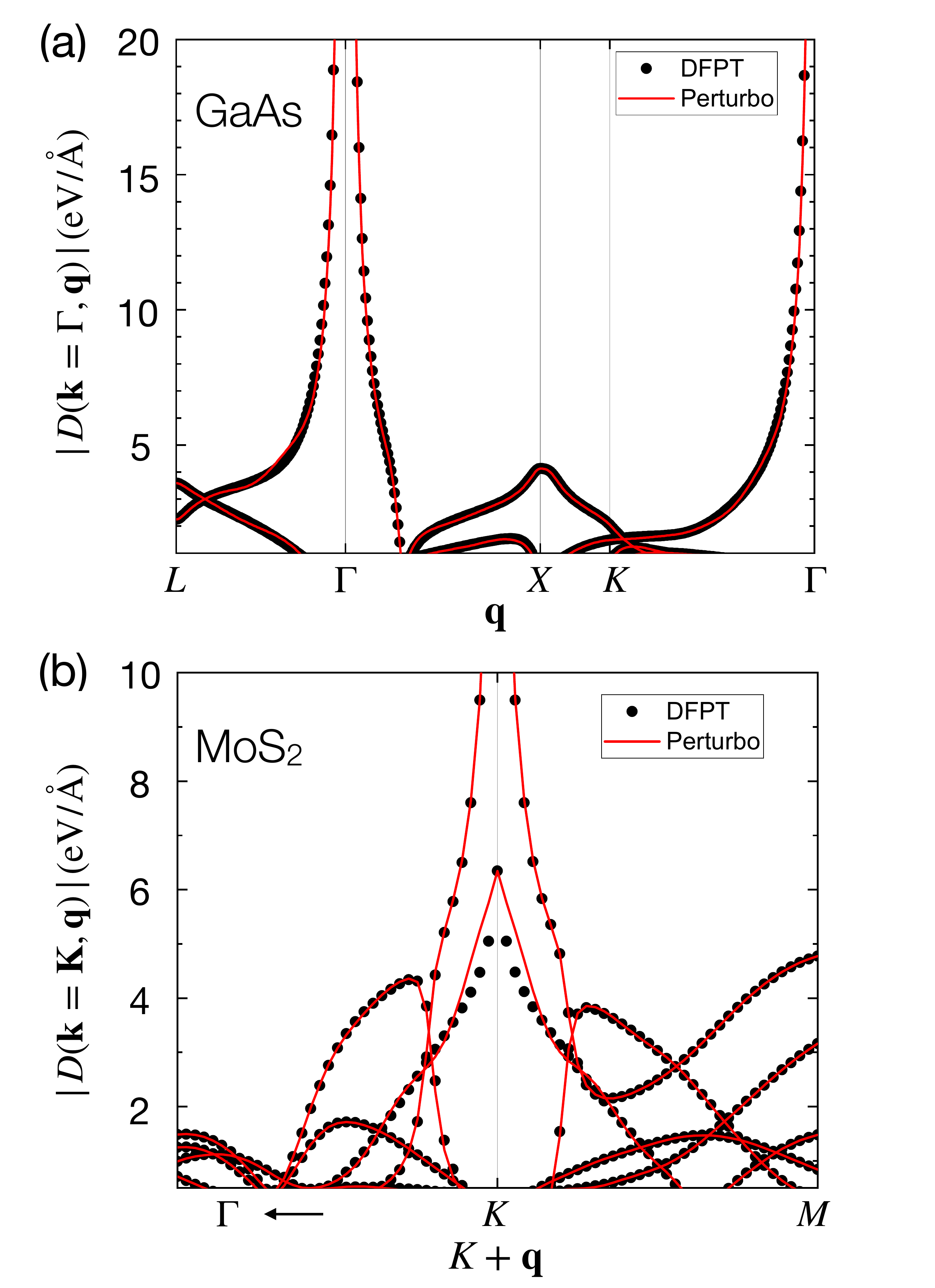}
\caption{ The $e$-ph deformation potential [see Eq.~(\ref{eq:def_pot})] in (a) GaAs and (b) monolayer MoS$_{2}$ computed using {\sc Perturbo} and compared with direct DFPT calculations for benchmarking.}
\label{fig:eph}
\end{figure}
\begin{figure}[!htb]
\centering
\includegraphics[width=0.88\columnwidth]{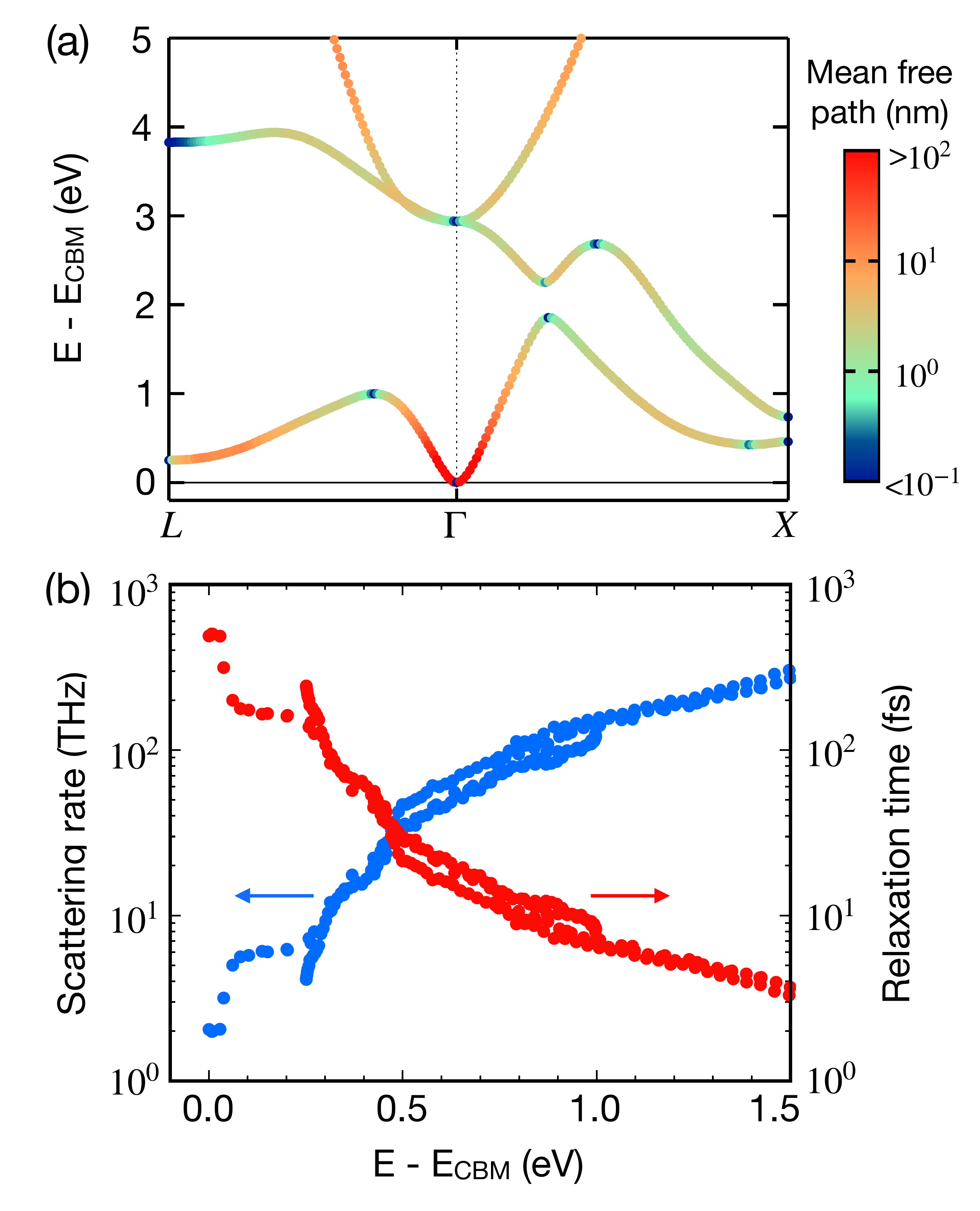}
\caption{(a) Band structure of GaAs overlaid with a log-scale color map of the electron mean free paths computed at 300~K. (b) The $e$-ph scattering rates and their inverse, the relaxation times, for the electronic states in (a) given as a function of electron energy. The energy zero is the CBM.}
\label{fig:mfp}
\end{figure}
\section{Examples} \label{sec:example}
In this section, we demonstrate the capabilities of the {\sc Perturbo} code with a few representative examples, including calculations on a polar bulk material (GaAs), a 2D material with SOC (monolayer MoS$_{2}$), and an organic crystal (naphthalene) with a relatively large unit cell with 36 atoms. \\
\indent
The ground state and band structure are computed using DFT with a plane-wave basis with the QE code; this is a preliminary step for all {\sc Perturbo} calculations, as discussed above. For GaAs, we use the same computational settings as in Ref.~\cite{Zhou2016}, namely a lattice constant of 5.556~\AA\  and a plane-wave kinetic energy cutoff of 72 Ry. For monolayer MoS$_{2}$, we use a 72 Ry kinetic energy cutoff, an experimental lattice constant of 3.16~\AA\ and a layer-normal vacuum spacing of 17~\AA. All calculations are carried out in the local density approximation of DFT using norm-conserving pseudopotentials. For MoS$_2$, we use fully relativistic norm-conserving pseudopotentials from Pseudo Dojo~\cite{Setten2018} to include SOC effects. \\
\begin{figure*}[!htb]
\centering
\includegraphics[width=2.0\columnwidth]{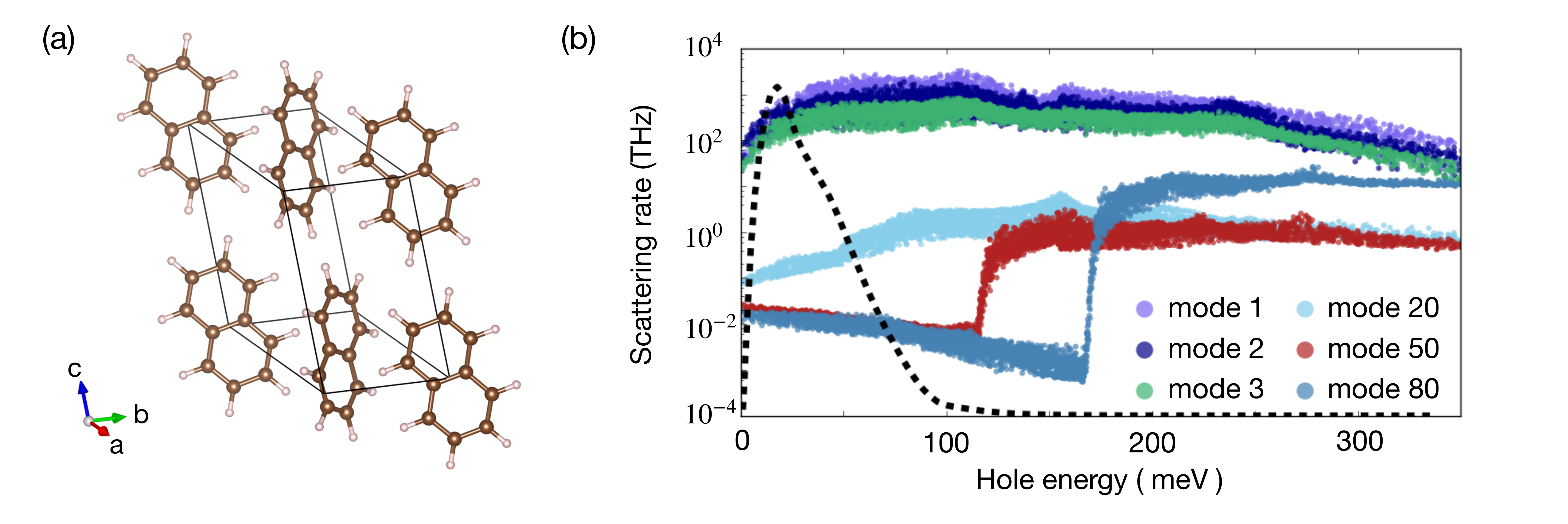}
\caption{
(a) Crystal structure of naphthalene, a prototypical organic molecular semiconductor with 36 atoms in the unit cell. (b) Mode-resolved $e$-ph scattering rates in naphthalene, computed at 300~K. Results are shown for three acoustic modes (1-3) associated with inter-molecular vibrations and three optical modes (20, 50, 80) associated with intra-molecular vibrations. The black dashed line represents the conductivity integrand in Eq.~(\ref{eq:cond_tdf}) and quantifies the relative contribution to transport as a function of electron energy. It is seen that only states within about 100~meV of the valence band maximum contribute to hole transport.}
\label{fig:naph}
\end{figure*}
\indent
Lattice dynamical properties and the $e$-ph perturbation potential [see Eq.~(\ref{eq:dvscf})] are computed on coarse $\bm{q}$-point grids of $8 \times 8 \times 8$ (GaAs) and 
$24 \times 24 \times 1$ (MoS$_{2}$) using DFPT as implemented in the QE code. 
We use the 2D Coulomb cutoff approach in DFPT for MoS$_{2}$ to remove the spurious interactions between layers~\cite{Sohier2016}. 
For naphthalene, we use the same computational settings as in Ref.~\cite{Lee2018} for the DFT and DFPT calculations. Note that we only perform the DFPT calculations for the irreducible $\bm{q}$-points in the BZ grid, following which we extend the dynamical matrices to the entire BZ grid using space group and time reversal symmetry [see Eq.~(\ref{eq:rot_dvscf})] with our \texttt{qe2pert.x} routines. \\
\indent
The {\sc Wannier90} code is employed to obtain localized Wannier functions in each material. Only the centers of the Wannier functions and the rotation matrices $\mathcal{U}$ [see Eq.~(\ref{eq:wannier_u})] are read as input by \texttt{qe2pert.x}. The Wannier functions for GaAs are constructed using a coarse $8 \times 8 \times 8$ $\bm{k}$-point grid and $sp^{3}$ orbitals centered at the Ga and As atoms as an initial guess. For MoS$_{2}$, we construct 22 Wannier functions using a coarse $24 \times 24 \times 1$ $\bm{k}$-point grid and an initial guess of $d$ orbitals on Mo and $s$ orbitals on S atoms, using both spin up and spin down orbitals. For naphthalene, we construct two Wannier functions for the two highest valence bands using a coarse $\bm{k}$-point grid of $4 \times 4 \times 4$. The selected columns of density matrix (SCDM) approach~\cite{Damle2017} is employed to generate an initial guess automatically.\\
\indent
Following the workflow in Fig.~\ref{fig:workflow}, we compute the $e$-ph matrix elements on the coarse $\bm{k}$- and $\bm{q}$-point grids and obtain the $e$-ph matrix elements in the localized Wannier basis using \texttt{qe2pert.x}.
%
% capabilities of our code.
\subsection{Interpolation of the $e$-ph matrix elements}
It is important to check the accuracy of all interpolations before performing transport and carrier dynamics calculations.  {\sc Perturbo} provides routines and calculation modes to compute and output the interpolated band structure, phonon dispersion, and $e$-ph matrix elements, which should be compared to the same quantities computed using DFT and DFPT to verify that the interpolation has worked as expected. 
The comparison is straightforward for the band structure and phonon dispersion. Here we focus on validating the interpolation of the $e$-ph matrix elements, a point often overlooked in $e$-ph calculations.
Since the $e$-ph matrix elements are gauge dependent, a direct comparison of 
$g_{mn\nu}\left(\bm{k},\bm{q}\right)$ from {\sc Perturbo} and DFPT calculations is not meaningful. Instead, we compute the absolute value of the $e$-ph matrix elements, $\left|g_{\nu}\left(\bm{k},\bm{q}\right)\right|$, in the gauge-invariant form of Eq.~(\ref{eq:g_abs}), and the closely related ``deformation potential'', which following Ref.~\cite{Sjakste2015} is defined as
\begin{equation}
D_{\nu}\left(\bm{k},\bm{q}\right) = \sqrt{2 \omega_{\nu\bm{q}}M_{tot}} \left|g_{\nu}\left(\bm{k},\bm{q}\right)\right|/\hbar,
\label{eq:def_pot}
\end{equation}
where $M_{tot}$ is the total mass of the atoms in the unit cell.  \\
\indent
We show in Fig.~\ref{fig:eph} the deformation potentials for the bulk polar material GaAs and the 2D polar material MoS$_{2}$. 
For GaAs, we choose the $\Gamma$-point as the initial electron momentum $\bm{k}$, and vary the phonon wave vector $\bm{q}$ along a high-symmetry path, computing $D_{\nu}\left(\bm{k},\bm{q}\right)$ for the highest valence band. For MoS$_{2}$, we choose the $K$-point as the initial electron momentum and vary the final electron momentum $K+\bm{q}$ along a high-symmetry path, computing $D_{\nu}\left(\bm{k},\bm{q}\right)$ by summing over the two lowest conduction bands. In both cases, results are computed for all phonon modes. The respective polar corrections are included in both materials, which is crucial to accurately interpolate the $e$-ph matrix elements at small $\bm{q}$. 
The results show clearly that the $e$-ph interpolation works as expected, giving interpolated matrix elements in close agreement with direct DFPT calculations.
%
%[basic check: DFT/DFPT vs interpolation]
%–	(Wannier interpolation vs DFPT (PH) )
%–	GaAs (3D polar correction),  MoS2 (2D polar correction)
%
%
\begin{figure*}[!htb]
\includegraphics[width=2.0\columnwidth]{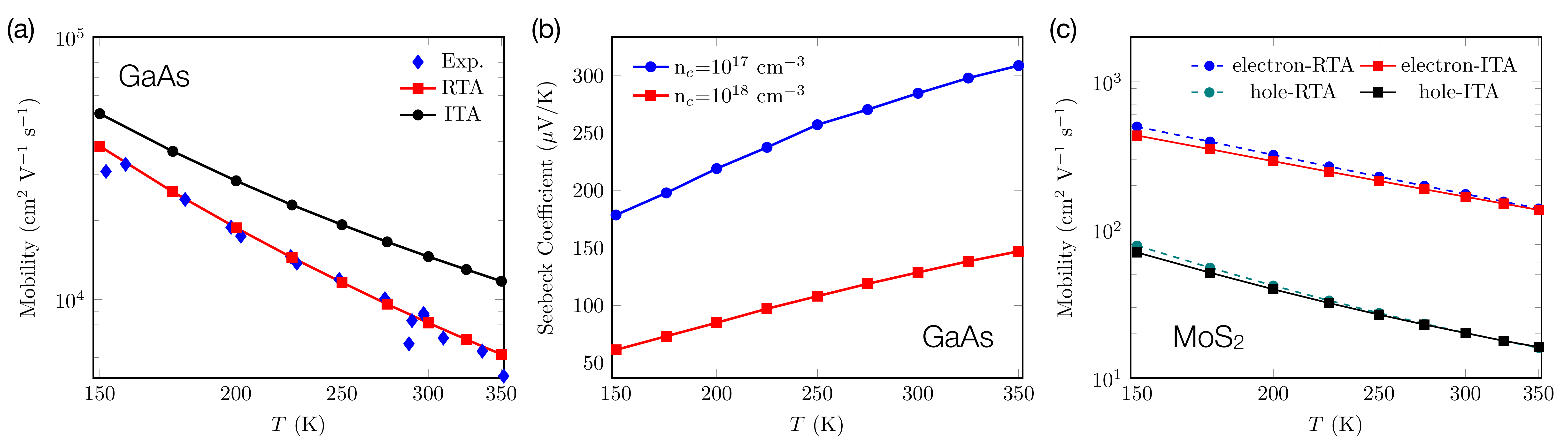}
\caption{(a) Electron mobility in GaAs as a function of temperature, computed using the RTA and ITA methods and compared with experimental data~\cite{Blood1972,Nichols1980}. (b) Temperature dependent Seebeck coefficient in GaAs, computed for two different carrier concentrations. (c) Electron and hole mobilities in monolayer MoS$_{2}$ as a function of temperature, computed using the RTA and ITA.}
\label{fig:trans}
\end{figure*}
\subsection{Scattering rates and electron mean free paths}\label{sec:scatter}
The \texttt{perturbo.x} routines can compute the $e$-ph scattering rate ($\Gamma_{n\bm{k}}$), relaxation time ($\tau_{n\bm{k}} = \Gamma_{n\bm{k}}^{-1}$), and electron mean free path ($L_{n\bm{k}} = \tau_{n\bm{k}} v_{n\bm{k}}$) for electronic states with any desired band and crystal momentum $\bm{k}$. Figure~\ref{fig:mfp} shows  the results for GaAs, in which we compute these quantities for the lowest few conduction bands and for $\bm{k}$-points along a high-symmetry path. \\
\indent
Carefully converging the scattering rate in Eq.~(\ref{eq:scatter_rate}) is important, but it can be far from trivial since in some cases convergence requires sampling millions of $\bm{q}$-points in the BZ.
A widely applicable scheme is to sample random $\bm{q}$-points uniformly, and compute the scattering rates for increasing numbers of $\bm{q}$-points until the scattering rates are converged.
%change in the scattering rates is within the convergence threshold.
However, uniform sampling can be nonoptimal in specific scenarios in which importance sampling can be used to speed up the convergence.
For polar materials such as GaAs, the scattering rate for electronic states near the conduction band minimum (CBM) is dominated by small-$\bm{q}$ (intravalley) LO phonon scattering. In this case, importance sampling with a Cauchy distribution~\cite{Zhou2016} that more extensively samples small $\bm{q}$ values can achieve convergence more effectively than uniform sampling. On the other hand, for electronic states in GaAs farther in energy from the band edges, the contributions to scattering are comparable for all phonon modes and momenta, so uniform sampling is effective because it avoids bias in the sampling. To treat optimally both sets of electronic states in polar materials like GaAs, 
%all the electronic states on the same footing and also reduce computational cost at the same time, 
we implement a polar split scheme in {\sc Perturbo}, which is detailed in Ref.~\cite{Zhou2016}. 
The scattering rates shown in Fig~\ref{fig:mfp}(b) are obtained with this approach.\\
%
%
%sampling random $\bm{q}$ points from a Cauchy distribution could achieve convergence with less $\bm{q}$ points comparing to uniform sampling. However, for high energy states, the contributions to scattering  from different phonon modes and momentum are comparable, uniform sampling is required to avoid biased sampling. 
%In order to treat all the electronic states on the same footing and reduce computational cost at the same time, a polar split scheme was introduced for polar materials, which is detailed in Ref.~\cite{Zhou2016}. The results in Fig.~\ref{fig:mfp} are computed using the polar split scheme. \\
%
%–	GaAs (electron)
%–	Naphetane, (SCDM)  (hole)
\indent
\hspace{3pt}To analyze the dominant scattering mechanism for charge transport, it is useful to resolve the contributions to the total scattering rate from different phonon modes. Figure~\ref{fig:naph} shows the mode-resolved scattering rates for holes in naphthalene computed at 300~K~\cite{Lee2018}. 
A unit cell of naphthalene includes two molecules, for a total of 36 atoms [see Fig.~\ref{fig:naph}(a)] and 108 phonon modes. Figure~\ref{fig:naph}(b) shows the contribution to the scattering rate from the three acoustic phonon modes (modes 1$-$3), which are associated with inter-molecular vibrations. The contributions from three optical modes (modes 20, 50, 80) associated with intra-molecular atomic vibrations are also shown. In the relevant energy range for transport, as shown by the dashed line in Fig.~\ref{fig:naph}(b), the inter-molecular modes dominate hole carrier scattering. This analysis is simple to carry out in {\sc Perturbo} because the code can output the TDF and mode-resolved scattering rates.
\begin{figure*}[!htb]
\centering
\includegraphics[width=1.9\columnwidth]{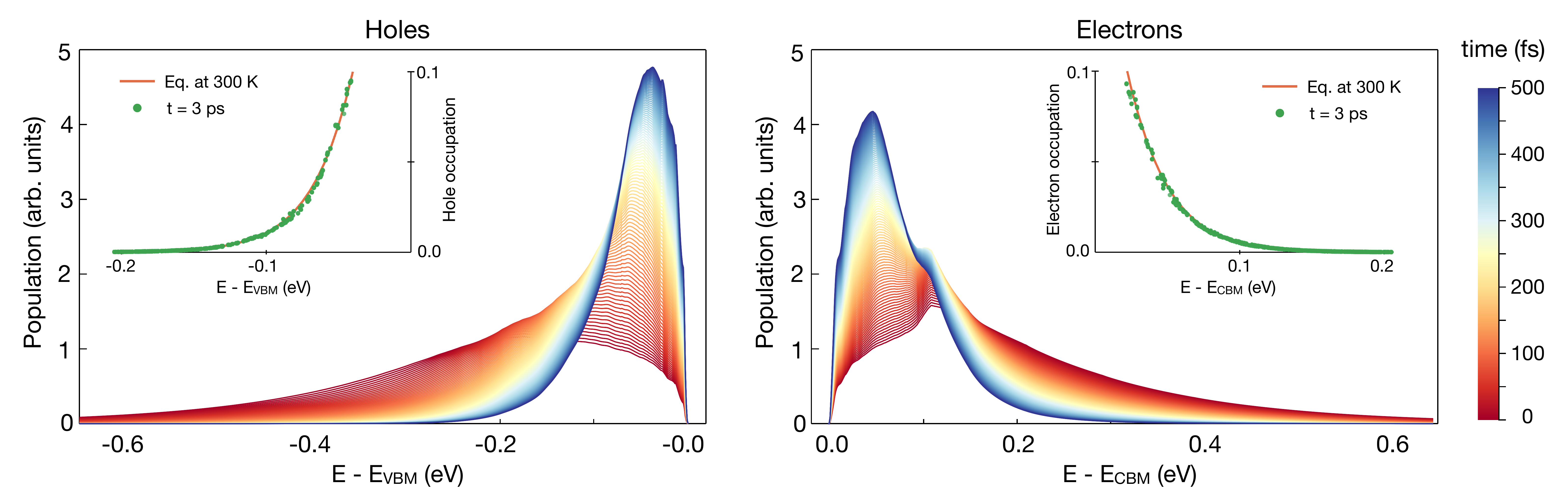}
\caption{
Hot carrier cooling simulation in silicon, for holes (left panel) and electrons (right panel). The time evolution of the carrier population $p(E, t)$ in Eq.~(\ref{eq:pop}) is shown, with the simulation time $t$ color-coded. The carrier occupations $f_{n\bm{k}}(t)$ at $t$ = 3~ps are compared in the inset with the equilibrium Fermi-Dirac distribution at 300~K (red curve) for the same carrier concentration. }
\label{fig:cdyna}
\end{figure*}
\subsection{Charge transport}
We present charge transport calculations for GaAs and monolayer MoS$_2$ as examples of a bulk and 2D material, respectively. Since both materials are polar, the polar corrections to the $e$-ph matrix elements are essential, as shown in Fig.~\ref{fig:eph}.
For MoS$_{2}$, we include SOC in the calculation since it plays an important role. In MoS$_2$, SOC splits the states near the valence band edge, so its inclusion has a significant impact on the $e$-ph scattering rates and mobility for hole carriers. 
The transport properties can be computed in {\sc Perturbo} either within the RTA or with the iterative solution of the linearized BTE. In the RTA, one can use pre-computed state-dependent scattering rates (see Sec.~\ref{sec:scatter}) or compute the scattering rates on the fly during the transport calculation using regular MP $\bm{k}$- and $\bm{q}$-point grids. In some materials, the RTA is inadequate and the more accurate iterative solution of the BTE in Eq.~(\ref{eq:BTE_iter}) is required. {\sc Perturbo} implements an efficiently parallelized routine to solve the BTE iteratively (see Sec.~\ref{sec:parallel}). \\
\indent
Figure~\ref{fig:trans}(a) shows the computed electron mobility in GaAs as a function of temperature and compares results obtained with the RTA and the iterative approach (ITA). 
The calculation uses a carrier concentration of $n_{c}\!=\!10^{17}$ cm$^{-3}$; we have checked that the mobility is almost independent of $n_{c}$ below 10$^{18}$ cm$^{-3}$. We perform the calculation using an ultra-fine BZ grid of $600 \times 600 \times 600$ for both $\bm{k}$- and $\bm{q}$-points, and employ a small (5 meV) Gaussian broadening to approximate the $\delta$ function in the $e$-ph scattering terms in Eq.~(\ref{eq:BTE_W}). To reduce the computational cost, we select a 200 meV energy window near the CBM, which covers the entire energy range relevant for transport. In GaAs, the ITA gives a higher electron mobility than the RTA, as shown in Fig.~\ref{fig:trans}(a), and this discrepancy increases with temperature. The temperature dependence of the electron mobility 
is also slightly different in the RTA and ITA results. The RTA calculation is in better agreement with experimental data~\cite{Blood1972,Nichols1980} than results from the more accurate ITA. 
Careful analysis, carried out elsewhere~\cite{Lee2019}, shows that correcting the band structure~\cite{Ma2018} to obtain an accurate effective mass and including two-phonon scattering processes~\cite{Lee2019} are both essential to obtain ITA results in agreement with experiment. \\
\indent
\hspace{-8pt}The Seebeck coefficient can be computed at negligible cost as a post-processing step of the mobility calculation. Figure~\ref{fig:trans}(b) shows the temperature dependent Seebeck coefficient in GaAs, computed using the ITA at two different carrier concentrations.  The computed value at 300~K and $n_{c}\!=\!10^{18}$ cm$^{-3}$ is about 130 $\mu$V/K, in agreement with the experimental value of $\sim$150 $\mu$V/K~\cite{Madelung2002}. 
Our results also show that the Seebeck coefficient increases 
with temperature and for decreasing carrier concentrations (here, we tested $n_{c}\!= \!10^{17}$~cm$^{-3}$), consistent with experimental data near room temperature~\cite{Madelung2002,Homm2008}. 
Note that the phonon occupations are kept fixed at their equilibrium value in our calculations, so the phonon drag effect, which is particularly important at low temperature, is neglected, and only the diffusive contribution to the Seebeck coefficient is computed. 
To include phonon drag one needs to include nonequilibrium phonon effects in the Seebeck coefficient calculation. 
%consider the coupled dynamics of electrons and phonons and include effects due to the nonequilibrium phonons 
%the contribution to the thermopower of the nonequilibrium phonons. Fully
While investigating the coupled dynamics of electrons and phonons remains an open challenge in first-principles calculations~\cite{Fiorentini2016,Protik2019}, we are developing an approach to time-step the coupled electron and phonon BTEs, and plan to make it available in a future version of {\sc Perturbo}. \\
%
%is under development in the  code, and will be released in a future version. 
%setup grid, energy windows
%RTA,  
% iterative, restart. 
%–	Grid Setup. 
%–	GaAs, electron, (RTA, iterative) (discuss Nien-En’s two phonon paper)
%–	MoS2 (2D) (SOC)
\indent
Figure~\ref{fig:trans}(c) shows the electron and hole mobilities in monolayer MoS$_{2}$, computed for a carrier concentration of $2 \times 10^{12}$ cm$^{-2}$ and for temperatures between 150$-$350~K. A fine BZ grid of $180 \times 180 \times 1$ is employed for both $\bm{k}$- and $\bm{q}$-points. 
Different from GaAs, the RTA and ITA give very close results for both the electron and hole mobilities, with only small differences at low temperature. The computed electron mobility at room temperature is about 168~cm$^{2}$/V$\,$s, in agreement with the experimental value of 150~cm$^{2}$/V$\,$s~\cite{Yu2016}; the computed hole mobility at 300~K is about 20~cm$^{2}$/V$\,$s. 
%
%3, electron, hole mobility, seebeack coef.:
%
%2, scattering rate and electron mean free path
%
\subsection{Ultrafast dynamics} \label{sec:example_silicon}
We demonstrate nonequilibrium ultrafast dynamics simulations using hot carrier cooling in silicon as an example. Excited (so-called “hot”) carriers can be generated in a number of ways in semiconductors, including injection from a contact or excitation with light. In a typical scenario, the excited carriers relax to the respective band edge in a sub-picosecond time scale by emitting phonons. While ultrafast carrier dynamics has been investigated extensively in experiments, for example with ultrafast optical spectroscopy, first-principles calculations of ultrafast dynamics, and in particular of hot carriers in the presence of $e$-ph interactions, are a recent development~\cite{Bernardi2014,Jhalani2017,Sadasivam2017,Molina2017}. \\
\indent
We perform a first-principles simulation of hot carrier cooling in silicon. We use a lattice constant of 5.389 \AA\ for DFT and DFPT calculations, which are carried out within the local density approximation and with norm-conserving pseudopotentials. Following the workflow in Fig.~\ref{fig:workflow}, we construct 8 Wannier functions using 
$sp^{3}$ orbitals on Si atoms as an initial guess, and obtain the $e$-ph matrix elements in the Wannier basis using coarse $8 \times 8 \times 8$ $\bm{k}$- and $\bm{q}$-point grids. \\
\begin{figure*}[!htb]
\centering
\includegraphics[width=1.8\columnwidth]{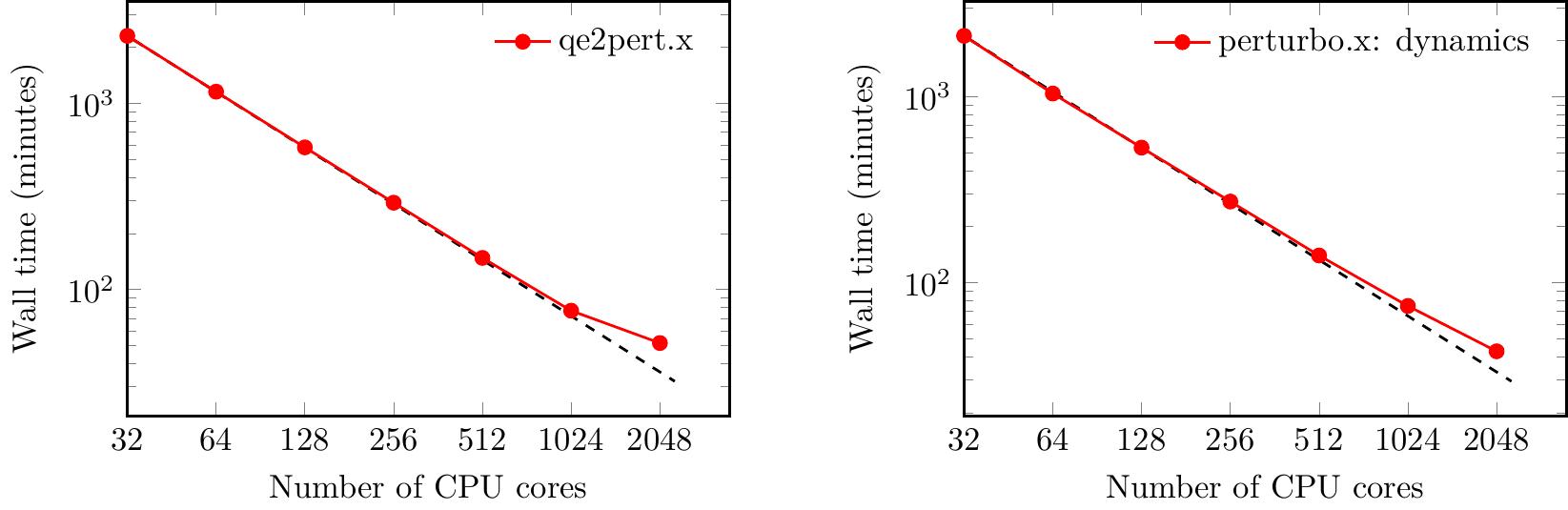}
\caption{
Performance scaling with the number of CPU cores of \texttt{qe2pert.x} and \texttt{perturbo.x}. The black dashed line indicates the ideal linear scaling.  For \texttt{perturbo.x}, we show the performance of the carrier dynamics simulation, which is the most time-consuming task of \texttt{perturbo.x}.}
\label{fig:scale}
\end{figure*}
\indent 
\hspace{-7pt}Starting from the $e$-ph matrix elements in the Wannier basis, the electron Hamiltonian, and interatomic force constants computed by \texttt{qe2pert.x}, 
we carry out hot carrier cooling simulations, separately for electron and hole carriers,  using a fine $\bm{k}$-point grid of $100 \times 100 \times 100$ for the electrons (or holes) and a $50 \times 50 \times 50$ $\bm{q}$-point grid for the phonons. 
A small Gaussian broadening of 8 meV is employed to approximate the $\delta$ function in the $e$-ph scattering terms in Eq.~(\ref{eq:BTE_W}). The phonon occupations are kept fixed at their equilibrium value at 300~K. 
We set the initial electron and hole occupations to hot Fermi-Dirac distributions at 1500~K, each with a chemical potential corresponding to a carrier concentration of 10$^{19}$ cm$^{-3}$. We explicitly time-step the BTE in Eq.~(\ref{eq:BTE_E}) (with external electric field set to zero) using the RK4 method with a small time step of 1 fs, 
and obtain nonequilibrium carrier occupations $f_{n\bm{k}}(t)$ as a function of time $t$ up to 3 ps, for a total of 3,000 time steps. The calculation takes only about 2.5 hours using 128 CPU cores due to the efficient parallelization (see Sec.~\ref{sec:parallel}). 
To visualize the time evolution of the carrier distributions, we compute the BZ averaged energy-dependent carrier population, 
\begin{equation}
p(E, t) = \sum_{n\bm{k}} f_{n\bm{k}}(t) \delta(\varepsilon_{n\bm{k}} - E), 
\label{eq:pop}
\end{equation}
using the tetrahedron integration method. The carrier population $p(E, t)$ characterizes 
the time-dependent energy distribution of the carriers. \\
%
%workflow figures
%
\indent
Figure~\ref{fig:cdyna} shows the evolution of the electron and hole populations due to $e$-ph scattering. For both electrons and holes, the high-energy tail in the initial population decays rapidly as the carriers quickly relax toward energies closer to the band edge within 500 fs. The carrier concentration is conserved during the simulation, validating the accuracy of our RK4 implementation to time-step the BTE. 
The hole relaxation is slightly faster than the electron relaxation in silicon, as shown in Fig.~\ref{fig:cdyna}. 
The sub-picosecond time scale we find for the hot carrier cooling is consistent with experiment and with previous calculations using a simplified approach~\cite{Bernardi2014}. Although the hot carriers accumulate near the band edge in a very short time, it takes longer for the carriers to fully relax to a 300~K thermal equilibrium distribution, up to several picoseconds in our simulation. We show in the inset of Fig.~\ref{fig:cdyna} that the carrier occupations $f_{n\bm{k}}$ at 3~ps reach the correct long-time limit for our simulation, namely an equilibrium Fermi-Dirac distribution at 
the 300~K lattice temperature. 
Since we keep the phonon occupations fixed and neglect phonon-phonon scattering, hot-phonon effects are ignored in our simulation. 
Similarly, electron-electron scattering, which is typically important at high carrier concentrations or in metals, is also not included in the current version of the code. Electron-electron scattering and coupled electron and phonon dynamics (including phonon-phonon collisions) are both under development.
%
%Figure~\ref{fig:cdyna} shows the hot cooling simulation on silicon. 
%-- Silicon, hot carrier cooling
%- [method GaN work and cite]
%1: g interpolation: 3D and 2D polar correction. (GaAs, MoS23)
%4, ultrafast carrier dynamics;
% 1) silicon
% 2) MoS2. 
%
%
\section{Parallelization and performance} \label{sec:parallel}
Transport calculations and ultrafast dynamics simulations on large systems can be computationally demanding and require a large amount of memory. Efficient parallelization is critical to run these calculations on high-performance computing (HPC) systems.  To fully take advantage of the typical HPC architecture, we implement a hybrid MPI and OpenMP parallelization, which combines distributed memory parallelization among different nodes using MPI and 
on-node shared memory parallelization using OpenMP. \\
\indent
For small systems, we observe a similar performance by running {\sc Perturbo} on a single node in either pure MPI or 
pure OpenMP modes, or in hybrid MPI plus OpenMP mode. 
However, for larger systems with several atoms to tens of atoms in the unit cell, running on multiple nodes with MPI plus OpenMP parallelization significantly improves the performance and leads to a better scaling with CPU core number. 
Compared to pure MPI, the hybrid MPI plus OpenMP scheme reduces communication needs and memory consumption, and improves load balance. We demonstrate the efficiency of our parallelization strategy using two examples: computing $e$-ph matrix elements on coarse grids, which is the most time-consuming task of \texttt{qe2pert.x}, and simulating nonequilibrium carrier dynamics, the most time-consuming task of \texttt{perturbo.x}.  \\
\indent
For the calculation of $e$-ph matrix elements on coarse grids [see Eq.~(\ref{eq:eph_mat_bare2})], our implementation uses MPI parallelization for $\bm{q}_{c}$-points and OpenMP parallelization for $\bm{k}_{c}$-points. We distribute the $\bm{q}_{c}$-points among the MPI processes, so that each MPI process either reads the self-consistent perturbation potential from file (if $\bm{q}_{c}$ is in the irreducible wedge) or computes it from the perturbation potential of the corresponding irreducible point using Eq.~(\ref{eq:rot_dvscf}). Each MPI process then computes $g(\bm{k}_{c}, \bm{q}_{c})$ for all the $\bm{k}_{c}$-points with OpenMP parallelization. \\
\indent
For carrier dynamics simulations, the $\bm{k}$-points on the fine grid are distributed among MPI processes to achieve an optimal load balance. 
Each process finds all the possible scattering channels involving its subset of $\bm{k}$-points, and stores these scattering processes locally as $(\bm{k}, \bm{q})$ pairs. 
%
%Each process keeps a subset of $\bm{k}$ points that are distributed to it, finds all the possible scattering processes involves $\bm{k}$ of its subset, and stores these scattering channels as $(\bm{k}, \bm{q})$ pairs.  Note that the $(\bm{k}, \bm{q})$ pairs are local to each processes. 
One can equivalently think of this approach as distributing the $(\bm{k}, \bm{q})$ pairs for the full $\bm{k}$-point set among MPI processes.
After this step, we use OpenMP parallelization over local $(\bm{k}, \bm{q})$ pairs to compute the $e$-ph matrix elements and perform the scattering integral.
The iterative BTE in Eq.~(\ref{eq:BTE_iter}) is also parallelized with the same approach.
This parallelization scheme requires minimum communication among MPI processes; for example, only one MPI reduction operation is required to collect the contributions from different processes for the integration in Eq.~(\ref{eq:BTE_iter}). \\
%\subsubsection{performance}
\indent
We test the parallelization performance of \texttt{qe2pert.x} and \texttt{perturbo.x} using calculations on MoS$_{2}$ and silicon as examples (see Sec.~\ref{sec:example} for the computational details). We run the carrier dynamics simulation on silicon using the Euler method with a time step of 1~fs and a total simulation time of 50~ps, for a total of 50,000 steps.
The tests are performed on the Cori system of the National Energy Research Scientific Computing center (NERSC). 
We use the Intel Xeon ``Haswell'' processor nodes, where each node has 32 CPU cores with a clock frequency of 2.3 GHz. \\
\indent
Figure~\ref{fig:scale} shows the wall time of the test calculations using different numbers of CPU cores. 
Both \texttt{qe2pert.x} and \texttt{perturbo.x} show rather remarkable scaling that is close to the ideal linear-scaling limit up to 1,024 CPU cores.
The scaling of \texttt{qe2pert.x} for 2,048 CPU cores is less than ideal mainly because each MPI process has an insufficient computational workload, so that communication and I/O become a bottleneck. 
The current release uses the serial HDF5 library, so after computing $g(\bm{k}_{c}, \bm{q}_{c})$, the root MPI process needs to collect $g(\bm{k}_{c}, \bm{q}_{c})$ from different processes and write them to disk (or vice versa when loading data), which is a serial task that costs $\sim$14\% of the total wall time when using 2048 CPU cores. Parallel I/O using the parallel HDF5 library could further improve the performance and overall scaling in this massively parallel example. We plan to work on this and related improvements to I/O in future releases. 
%
%For example, computing $g(\bm{k}_{c}, \bm{q}_{c})$ takes about $\sim$44 minutes, while collecting $g(\bm{k}_{c}, \bm{q}_{c})$ from different processes and writing them to disk, a task that is not parallelized, takes roughly $\sim$7 minutes. 
%The current release of {\sc Perturbo} uses serial HDF5 library, so only the root MPI process reads data from disk and distributes them to other processes (or vice versa). Parallel I/O using the parallel HDF5 library could further improve the performance and overall scaling in this massively parallel example, and we plan to work on this and related improvements to I/O in future releases.
%
%
\section{Conclusions and outlook} \label{sec:conc}
In conclusion, we present our software, {\sc Perturbo}, for first-principles calculations of charge transport properties and simulation of ultrafast carrier dynamics in bulk and 2D materials. 
The software contains an interface program to read results from DFT and DFPT calculations from the QE code. 
The core program of  {\sc Perturbo} performs various computational tasks, such as computing $e$-ph scattering rates, electron mean free paths, electrical conductivity, mobility, and the Seebeck coefficient. The code can also simulate the nonequilibrium dynamics of excited electrons. 
Wannier interpolation and symmetry are employed to greatly reduce the computational cost. 
SOC and the polar corrections for bulk and 2D materials are supported and have been carefully tested. 
We demonstrate these features with representative examples. We also show the highly promising scaling of {\sc Perturbo} on massively parallel HPC architectures, owing to its effective implementation of hybrid MPI plus OpenMP parallelization.\\
\indent
Several features are currently under development for the next major release, including spin-phonon relaxation times and spin dynamics~\cite{Park2020}, transport calculations in the large polaron regime using the Kubo formalism~\cite{Zhou2019}, and charge transport and ultrafast dynamics for coupled electrons and phonons, among others. As an alternative to Wannier functions, interpolation using atomic orbitals~\cite{Agapito2018} will also be supported in a future release. We will extend the interface program to support additional external codes, such as the TDEP software for temperature-dependent lattice dynamical calculations~\cite{Zhou2018,Hellman2013}.

\section*{Acknowledgments}
We thank V.A. Jhalani and B.K. Chang for fruitful discussions. This work was supported by the National Science Foundation under Grants No.~ACI-1642443 for code development and CAREER-1750613 for theory development.
J.-J.Z. acknowledges support by the Joint Center for Artificial Photosynthesis, a DOE Energy Innovation Hub, supported through the Office of Science of the U.S. Department of Energy under Award No.~DE-SC0004993. 
J.P. acknowledges support by the Korea Foundation for Advanced Studies. 
I-T.L. was supported by the Air Force Office of Scientific Research through the Young Investigator Program, Grant FA9550-18-1-0280. X.T. was supported by the Resnick Institute at Caltech.
This research used resources of the National Energy Research Scientific Computing Center, a DOE Office of Science User Facility supported by the Office of Science of the U.S. Department of Energy under Contract No. DE-AC02-05CH11231.

%\section*{References}
\bibliography{references}

\begin{thebibliography}{60}
\providecommand{\natexlab}[1]{#1}
\providecommand{\url}[1]{\texttt{#1}}
\providecommand{\urlprefix}{URL }
\expandafter\ifx\csname urlstyle\endcsname\relax
  \providecommand{\doi}[1]{doi:\discretionary{}{}{}#1}\else
  \providecommand{\doi}[1]{doi:\discretionary{}{}{}\begingroup
  \urlstyle{rm}\url{#1}\endgroup}\fi
\providecommand{\bibinfo}[2]{#2}

\bibitem[{Rossi and Kuhn(2002)}]{Rossi2002}
\bibinfo{author}{F.~Rossi}, \bibinfo{author}{T.~Kuhn}, \bibinfo{title}{Theory
  of ultrafast phenomena in photoexcited semiconductors},
  \bibinfo{journal}{Rev. Mod. Phys.} \bibinfo{volume}{74}~(\bibinfo{number}{3})
  (\bibinfo{year}{2002}) \bibinfo{pages}{895--950}.

\bibitem[{Ulbricht et~al.(2011)Ulbricht, Hendry, Shan, Heinz, and
  Bonn}]{Ulbricht2011}
\bibinfo{author}{R.~Ulbricht}, \bibinfo{author}{E.~Hendry},
  \bibinfo{author}{J.~Shan}, \bibinfo{author}{T.~F. Heinz},
  \bibinfo{author}{M.~Bonn}, \bibinfo{title}{Carrier dynamics in semiconductors
  studied with time-resolved terahertz spectroscopy}, \bibinfo{journal}{Rev.
  Mod. Phys.} \bibinfo{volume}{83}~(\bibinfo{number}{2}) (\bibinfo{year}{2011})
  \bibinfo{pages}{543--586}.

\bibitem[{Egger et~al.(2018)Egger, Bera, Cahen, Hodes, Kirchartz, Kronik
  et~al.}]{Egger2018}
\bibinfo{author}{D.~A. Egger}, \bibinfo{author}{A.~Bera},
  \bibinfo{author}{D.~Cahen}, \bibinfo{author}{G.~Hodes},
  \bibinfo{author}{T.~Kirchartz}, \bibinfo{author}{L.~Kronik}, et~al.,
  \bibinfo{title}{What remains unexplained about the properties of halide
  perovskites?}, \bibinfo{journal}{Adv. Mater.}
  \bibinfo{volume}{30}~(\bibinfo{number}{20}) (\bibinfo{year}{2018})
  \bibinfo{pages}{1800691}.

\bibitem[{Baroni et~al.(2001)Baroni, de~Gironcoli, Dal~Corso, and
  Giannozzi}]{Baroni2001}
\bibinfo{author}{S.~Baroni}, \bibinfo{author}{S.~de~Gironcoli},
  \bibinfo{author}{A.~Dal~Corso}, \bibinfo{author}{P.~Giannozzi},
  \bibinfo{title}{Phonons and related crystal properties from
  density-functional perturbation theory}, \bibinfo{journal}{Rev. Mod. Phys.}
  \bibinfo{volume}{73}~(\bibinfo{number}{2}) (\bibinfo{year}{2001})
  \bibinfo{pages}{515--562}.

\bibitem[{Giannozzi et~al.(2009)Giannozzi, Baroni, Bonini, Calandra, Car
  et~al.}]{Giannozzi2009}
\bibinfo{author}{P.~Giannozzi}, \bibinfo{author}{S.~Baroni},
  \bibinfo{author}{N.~Bonini}, \bibinfo{author}{M.~Calandra},
  \bibinfo{author}{R.~Car}, et~al., \bibinfo{title}{{QUANTUM} {ESPRESSO}: a
  modular and open-source software project for quantum simulations of
  materials}, \bibinfo{journal}{J. Phys.: Condens. Matter}
  \bibinfo{volume}{21}~(\bibinfo{number}{39}) (\bibinfo{year}{2009})
  \bibinfo{pages}{395502}.

\bibitem[{Giannozzi et~al.(2017)Giannozzi, Andreussi, Brumme, Bunau, Nardelli
  et~al.}]{Giannozzi2017}
\bibinfo{author}{P.~Giannozzi}, \bibinfo{author}{O.~Andreussi},
  \bibinfo{author}{T.~Brumme}, \bibinfo{author}{O.~Bunau},
  \bibinfo{author}{M.~B. Nardelli}, et~al., \bibinfo{title}{Advanced
  capabilities for materials modelling with {QUANTUM} {ESPRESSO}},
  \bibinfo{journal}{J. Phys.: Condens. Matter}
  \bibinfo{volume}{29}~(\bibinfo{number}{46}) (\bibinfo{year}{2017})
  \bibinfo{pages}{465901}.

\bibitem[{Ziman(2007)}]{Ziman2007}
\bibinfo{author}{J.~M. Ziman}, \bibinfo{title}{Electrons and phonons: {The}
  theory of transport phenomena in solids}, \bibinfo{publisher}{Oxford
  University Press}, \bibinfo{address}{New York}, \bibinfo{year}{2007}.

\bibitem[{Park et~al.(2020)Park, Zhou, and Bernardi}]{Park2020}
\bibinfo{author}{J.~Park}, \bibinfo{author}{J.-J. Zhou},
  \bibinfo{author}{M.~Bernardi}, \bibinfo{title}{Spin-phonon relaxation times
  in centrosymmetric materials from first principles}, \bibinfo{journal}{Phys.
  Rev. B} \bibinfo{volume}{101} (\bibinfo{year}{2020}) \bibinfo{pages}{045202}.

\bibitem[{Lu et~al.(2019{\natexlab{a}})Lu, Zhou, and Bernardi}]{Lu2019}
\bibinfo{author}{I.-T. Lu}, \bibinfo{author}{J.-J. Zhou},
  \bibinfo{author}{M.~Bernardi}, \bibinfo{title}{Efficient ab initio
  calculations of electron-defect scattering and defect-limited carrier
  mobility}, \bibinfo{journal}{Phys. Rev. Mater.} \bibinfo{volume}{3}
  (\bibinfo{year}{2019}{\natexlab{a}}) \bibinfo{pages}{033804}.

\bibitem[{Lu et~al.(2019{\natexlab{b}})Lu, Park, Zhou, and Bernardi}]{Lu2019a}
\bibinfo{author}{I.-T. Lu}, \bibinfo{author}{J.~Park}, \bibinfo{author}{J.-J.
  Zhou}, \bibinfo{author}{M.~Bernardi}, \bibinfo{title}{Ab initio
  electron-defect interactions using {Wannier} functions},
  \bibinfo{journal}{arXiv:1910.14516 [cond-mat.mtrl-sci]} .

\bibitem[{Chen et~al.(2019)Chen, Jhalani, Palummo, and Bernardi}]{Chen2019}
\bibinfo{author}{H.-Y. Chen}, \bibinfo{author}{V.~A. Jhalani},
  \bibinfo{author}{M.~Palummo}, \bibinfo{author}{M.~Bernardi},
  \bibinfo{title}{Ab initio calculations of exciton radiative lifetimes in bulk
  crystals, nanostructures, and molecules}, \bibinfo{journal}{Phys. Rev. B}
  \bibinfo{volume}{100} (\bibinfo{year}{2019}) \bibinfo{pages}{075135}.

\bibitem[{Zhou and Bernardi(2016)}]{Zhou2016}
\bibinfo{author}{J.-J. Zhou}, \bibinfo{author}{M.~Bernardi},
  \bibinfo{title}{\textit{Ab initio} electron mobility and polar phonon
  scattering in {GaAs}}, \bibinfo{journal}{Phys. Rev. B}
  \bibinfo{volume}{94}~(\bibinfo{number}{20}) (\bibinfo{year}{2016})
  \bibinfo{pages}{201201(R)}.

\bibitem[{Zhou et~al.(2018)Zhou, Hellman, and Bernardi}]{Zhou2018}
\bibinfo{author}{J.-J. Zhou}, \bibinfo{author}{O.~Hellman},
  \bibinfo{author}{M.~Bernardi}, \bibinfo{title}{Electron-phonon scattering in
  the presence of soft modes and electron mobility in ${\mathrm{SrTiO}}_{3}$
  perovskite from first principles}, \bibinfo{journal}{Phys. Rev. Lett.}
  \bibinfo{volume}{121}~(\bibinfo{number}{22}) (\bibinfo{year}{2018})
  \bibinfo{pages}{226603}.

\bibitem[{Jhalani et~al.(2017)Jhalani, Zhou, and Bernardi}]{Jhalani2017}
\bibinfo{author}{V.~A. Jhalani}, \bibinfo{author}{J.-J. Zhou},
  \bibinfo{author}{M.~Bernardi}, \bibinfo{title}{Ultrafast hot carrier dynamics
  in GaN and its impact on the efficiency droop}, \bibinfo{journal}{Nano Lett.}
  \bibinfo{volume}{17}~(\bibinfo{number}{8}) (\bibinfo{year}{2017})
  \bibinfo{pages}{5012--5019}.

\bibitem[{Lee et~al.(2018)Lee, Zhou, Agapito, and Bernardi}]{Lee2018}
\bibinfo{author}{N.-E. Lee}, \bibinfo{author}{J.-J. Zhou},
  \bibinfo{author}{L.~A. Agapito}, \bibinfo{author}{M.~Bernardi},
  \bibinfo{title}{Charge transport in organic molecular semiconductors from
  first principles: {The} bandlike hole mobility in a naphthalene crystal},
  \bibinfo{journal}{Phys. Rev. B} \bibinfo{volume}{97}~(\bibinfo{number}{11})
  (\bibinfo{year}{2018}) \bibinfo{pages}{115203}.

\bibitem[{Bernardi(2016)}]{Bernardi2016}
\bibinfo{author}{M.~Bernardi}, \bibinfo{title}{First-principles dynamics of
  electrons and phonons}, \bibinfo{journal}{Eur. Phys. J. B}
  \bibinfo{volume}{89}~(\bibinfo{number}{11}) (\bibinfo{year}{2016})
  \bibinfo{pages}{239}.

\bibitem[{Madsen and Singh(2006)}]{Madsen2006}
\bibinfo{author}{G.~K.~H. Madsen}, \bibinfo{author}{D.~J. Singh},
  \bibinfo{title}{{BoltzTraP}. {A} code for calculating band-structure
  dependent quantities}, \bibinfo{journal}{Comput. Phys. Commun.}
  \bibinfo{volume}{175}~(\bibinfo{number}{1}) (\bibinfo{year}{2006})
  \bibinfo{pages}{67--71}.

\bibitem[{Pizzi et~al.(2014)Pizzi, Volja, Kozinsky, Fornari, and
  Marzari}]{Pizzi2014}
\bibinfo{author}{G.~Pizzi}, \bibinfo{author}{D.~Volja},
  \bibinfo{author}{B.~Kozinsky}, \bibinfo{author}{M.~Fornari},
  \bibinfo{author}{N.~Marzari}, \bibinfo{title}{{BoltzWann}: {A} code for the
  evaluation of thermoelectric and electronic transport properties with a
  maximally-localized {Wannier} functions basis}, \bibinfo{journal}{Comput.
  Phys. Commun.} \bibinfo{volume}{185}~(\bibinfo{number}{1})
  (\bibinfo{year}{2014}) \bibinfo{pages}{422--429}.

\bibitem[{Bernardi et~al.(2014)Bernardi, Vigil-Fowler, Lischner, Neaton, and
  Louie}]{Bernardi2014}
\bibinfo{author}{M.~Bernardi}, \bibinfo{author}{D.~Vigil-Fowler},
  \bibinfo{author}{J.~Lischner}, \bibinfo{author}{J.~B. Neaton},
  \bibinfo{author}{S.~G. Louie}, \bibinfo{title}{\textit{Ab~initio} study of
  hot carriers in the first picosecond after sunlight absorption in silicon},
  \bibinfo{journal}{Phys. Rev. Lett.} \bibinfo{volume}{112}
  (\bibinfo{year}{2014}) \bibinfo{pages}{257402}.

\bibitem[{Mahan(2000)}]{Mahan2000}
\bibinfo{author}{G.~D. Mahan}, \bibinfo{title}{Many-particle physics},
  \bibinfo{publisher}{Springer US}, \bibinfo{edition}{3rd} edn.,
  \bibinfo{year}{2000}.

\bibitem[{Li(2015)}]{Li2015}
\bibinfo{author}{W.~Li}, \bibinfo{title}{Electrical transport limited by
  electron-phonon coupling from Boltzmann transport equation: An \textit{ab
  initio} study of Si, Al, and ${\mathrm{MoS}}_{2}$}, \bibinfo{journal}{Phys.
  Rev. B} \bibinfo{volume}{92}~(\bibinfo{number}{7}) (\bibinfo{year}{2015})
  \bibinfo{pages}{075405}.

\bibitem[{Fiorentini and Bonini(2016)}]{Fiorentini2016}
\bibinfo{author}{M.~Fiorentini}, \bibinfo{author}{N.~Bonini},
  \bibinfo{title}{Thermoelectric coefficients of $n$-doped silicon from first
  principles via the solution of the {Boltzmann} transport equation},
  \bibinfo{journal}{Phys. Rev. B} \bibinfo{volume}{94}~(\bibinfo{number}{8})
  (\bibinfo{year}{2016}) \bibinfo{pages}{085204}.

\bibitem[{Bl\"ochl et~al.(1994)Bl\"ochl, Jepsen, and Andersen}]{Bloechl1994}
\bibinfo{author}{P.~E. Bl\"ochl}, \bibinfo{author}{O.~Jepsen},
  \bibinfo{author}{O.~K. Andersen}, \bibinfo{title}{Improved tetrahedron method
  for {Brillouin}-zone integrations}, \bibinfo{journal}{Phys. Rev. B}
  \bibinfo{volume}{49}~(\bibinfo{number}{23}) (\bibinfo{year}{1994})
  \bibinfo{pages}{16223--16233}.

\bibitem[{Marzari et~al.(2012)Marzari, Mostofi, Yates, Souza, and
  Vanderbilt}]{Marzari2012}
\bibinfo{author}{N.~Marzari}, \bibinfo{author}{A.~A. Mostofi},
  \bibinfo{author}{J.~R. Yates}, \bibinfo{author}{I.~Souza},
  \bibinfo{author}{D.~Vanderbilt}, \bibinfo{title}{Maximally localized
  {Wannier} functions: {Theory} and applications}, \bibinfo{journal}{Rev. Mod.
  Phys.} \bibinfo{volume}{84}~(\bibinfo{number}{4}) (\bibinfo{year}{2012})
  \bibinfo{pages}{1419--1475}.

\bibitem[{Mostofi et~al.(2014)Mostofi, Yates, Pizzi, Lee, Souza, Vanderbilt,
  and Marzari}]{Mostofi2014}
\bibinfo{author}{A.~A. Mostofi}, \bibinfo{author}{J.~R. Yates},
  \bibinfo{author}{G.~Pizzi}, \bibinfo{author}{Y.-S. Lee},
  \bibinfo{author}{I.~Souza}, \bibinfo{author}{D.~Vanderbilt},
  \bibinfo{author}{N.~Marzari}, \bibinfo{title}{An updated version of
  wannier90: {A} tool for obtaining maximally-localised {Wannier} functions},
  \bibinfo{journal}{Comput. Phys. Commun.}
  \bibinfo{volume}{185}~(\bibinfo{number}{8}) (\bibinfo{year}{2014})
  \bibinfo{pages}{2309--2310}.

\bibitem[{Pizzi et~al.(2020)Pizzi, Vitale, Arita, Blügel, Freimuth, Géranton,
  Gibertini, Gresch, Johnson, Koretsune et~al.}]{Pizzi2020}
\bibinfo{author}{G.~Pizzi}, \bibinfo{author}{V.~Vitale},
  \bibinfo{author}{R.~Arita}, \bibinfo{author}{S.~Blügel},
  \bibinfo{author}{F.~Freimuth}, \bibinfo{author}{G.~Géranton},
  \bibinfo{author}{M.~Gibertini}, \bibinfo{author}{D.~Gresch},
  \bibinfo{author}{C.~Johnson}, \bibinfo{author}{T.~Koretsune}, et~al.,
  \bibinfo{title}{Wannier90 as a community code: new features and
  applications}, \bibinfo{journal}{J. Phys.: Condens. Matter}
  \bibinfo{volume}{32}~(\bibinfo{number}{16}) (\bibinfo{year}{2020})
  \bibinfo{pages}{165902}.

\bibitem[{Wang et~al.(2006)Wang, Yates, Souza, and Vanderbilt}]{Wang2006}
\bibinfo{author}{X.~Wang}, \bibinfo{author}{J.~R. Yates},
  \bibinfo{author}{I.~Souza}, \bibinfo{author}{D.~Vanderbilt},
  \bibinfo{title}{Ab initio calculation of the anomalous Hall conductivity by
  Wannier interpolation}, \bibinfo{journal}{Phys. Rev. B} \bibinfo{volume}{74}
  (\bibinfo{year}{2006}) \bibinfo{pages}{195118}.

\bibitem[{Souza et~al.(2001)Souza, Marzari, and Vanderbilt}]{Souza2001}
\bibinfo{author}{I.~Souza}, \bibinfo{author}{N.~Marzari},
  \bibinfo{author}{D.~Vanderbilt}, \bibinfo{title}{Maximally localized
  {Wannier} functions for entangled energy bands}, \bibinfo{journal}{Phys. Rev.
  B} \bibinfo{volume}{65}~(\bibinfo{number}{3}) (\bibinfo{year}{2001})
  \bibinfo{pages}{035109}.

\bibitem[{Yates et~al.(2007)Yates, Wang, Vanderbilt, and Souza}]{Yates2007}
\bibinfo{author}{J.~R. Yates}, \bibinfo{author}{X.~Wang},
  \bibinfo{author}{D.~Vanderbilt}, \bibinfo{author}{I.~Souza},
  \bibinfo{title}{Spectral and {Fermi} surface properties from {Wannier}
  interpolation}, \bibinfo{journal}{Phys. Rev. B}
  \bibinfo{volume}{75}~(\bibinfo{number}{19}) (\bibinfo{year}{2007})
  \bibinfo{pages}{195121}.

\bibitem[{Born and Huang(1954)}]{Born1954}
\bibinfo{author}{M.~Born}, \bibinfo{author}{K.~Huang},
  \bibinfo{title}{Dynamical theory of crystal lattices},
  \bibinfo{publisher}{Clarendon press}, \bibinfo{year}{1954}.

\bibitem[{Gonze and Lee(1997)}]{Gonze1997}
\bibinfo{author}{X.~Gonze}, \bibinfo{author}{C.~Lee}, \bibinfo{title}{Dynamical
  matrices, {Born} effective charges, dielectric permittivity tensors, and
  interatomic force constants from density-functional perturbation theory},
  \bibinfo{journal}{Phys. Rev. B} \bibinfo{volume}{55}~(\bibinfo{number}{16})
  (\bibinfo{year}{1997}) \bibinfo{pages}{10355--10368}.

\bibitem[{Giustino et~al.(2007)Giustino, Cohen, and Louie}]{Giustino2007}
\bibinfo{author}{F.~Giustino}, \bibinfo{author}{M.~L. Cohen},
  \bibinfo{author}{S.~G. Louie}, \bibinfo{title}{Electron-phonon interaction
  using {Wannier} functions}, \bibinfo{journal}{Phys. Rev. B}
  \bibinfo{volume}{76}~(\bibinfo{number}{16}) (\bibinfo{year}{2007})
  \bibinfo{pages}{165108}.

\bibitem[{Calandra et~al.(2010)Calandra, Profeta, and Mauri}]{Calandra2010}
\bibinfo{author}{M.~Calandra}, \bibinfo{author}{G.~Profeta},
  \bibinfo{author}{F.~Mauri}, \bibinfo{title}{Adiabatic and nonadiabatic phonon
  dispersion in a {Wannier} function approach}, \bibinfo{journal}{Phys. Rev. B}
  \bibinfo{volume}{82}~(\bibinfo{number}{16}) (\bibinfo{year}{2010})
  \bibinfo{pages}{165111}.

\bibitem[{Agapito and Bernardi(2018)}]{Agapito2018}
\bibinfo{author}{L.~A. Agapito}, \bibinfo{author}{M.~Bernardi},
  \bibinfo{title}{Ab initio electron-phonon interactions using atomic orbital
  wave functions}, \bibinfo{journal}{Phys. Rev. B}
  \bibinfo{volume}{97}~(\bibinfo{number}{23}) (\bibinfo{year}{2018})
  \bibinfo{pages}{235146}.

\bibitem[{Gonze et~al.(1994)Gonze, Charlier, Allan, and Teter}]{Gonze1994}
\bibinfo{author}{X.~Gonze}, \bibinfo{author}{J.-C. Charlier},
  \bibinfo{author}{D.~Allan}, \bibinfo{author}{M.~Teter},
  \bibinfo{title}{Interatomic force constants from first principles: The case
  of \ensuremath{\alpha}-quartz}, \bibinfo{journal}{Phys. Rev. B}
  \bibinfo{volume}{50} (\bibinfo{year}{1994}) \bibinfo{pages}{13035(R)}.

\bibitem[{Fröhlich(1954)}]{Frohlich1954}
\bibinfo{author}{H.~Fröhlich}, \bibinfo{title}{Electrons in lattice fields},
  \bibinfo{journal}{Adv. Phys.} \bibinfo{volume}{3}~(\bibinfo{number}{11})
  (\bibinfo{year}{1954}) \bibinfo{pages}{325--361}.

\bibitem[{Sjakste et~al.(2015)Sjakste, Vast, Calandra, and Mauri}]{Sjakste2015}
\bibinfo{author}{J.~Sjakste}, \bibinfo{author}{N.~Vast},
  \bibinfo{author}{M.~Calandra}, \bibinfo{author}{F.~Mauri},
  \bibinfo{title}{Wannier interpolation of the electron-phonon matrix elements
  in polar semiconductors: {Polar}-optical coupling in {GaAs}},
  \bibinfo{journal}{Phys. Rev. B} \bibinfo{volume}{92}~(\bibinfo{number}{5})
  (\bibinfo{year}{2015}) \bibinfo{pages}{054307}.

\bibitem[{Verdi and Giustino(2015)}]{Verdi2015}
\bibinfo{author}{C.~Verdi}, \bibinfo{author}{F.~Giustino},
  \bibinfo{title}{Fr\"ohlich electron-phonon vertex from first principles},
  \bibinfo{journal}{Phys. Rev. Lett.}
  \bibinfo{volume}{115}~(\bibinfo{number}{17}) (\bibinfo{year}{2015})
  \bibinfo{pages}{176401}.

\bibitem[{Sohier et~al.(2016)Sohier, Calandra, and Mauri}]{Sohier2016}
\bibinfo{author}{T.~Sohier}, \bibinfo{author}{M.~Calandra},
  \bibinfo{author}{F.~Mauri}, \bibinfo{title}{Two-dimensional Fr\"ohlich
  interaction in transition-metal dichalcogenide monolayers: Theoretical
  modeling and first-principles calculations}, \bibinfo{journal}{Phys. Rev. B}
  \bibinfo{volume}{94} (\bibinfo{year}{2016}) \bibinfo{pages}{085415}.

\bibitem[{Sohier et~al.(2017{\natexlab{a}})Sohier, Gibertini, Calandra, Mauri,
  and Marzari}]{Sohier2017}
\bibinfo{author}{T.~Sohier}, \bibinfo{author}{M.~Gibertini},
  \bibinfo{author}{M.~Calandra}, \bibinfo{author}{F.~Mauri},
  \bibinfo{author}{N.~Marzari}, \bibinfo{title}{Breakdown of optical phonons’
  splitting in two-dimensional materials}, \bibinfo{journal}{Nano Lett.}
  \bibinfo{volume}{17}~(\bibinfo{number}{6})
  (\bibinfo{year}{2017}{\natexlab{a}}) \bibinfo{pages}{3758--3763}.

\bibitem[{Sohier et~al.(2017{\natexlab{b}})Sohier, Calandra, and
  Mauri}]{Sohier2017a}
\bibinfo{author}{T.~Sohier}, \bibinfo{author}{M.~Calandra},
  \bibinfo{author}{F.~Mauri}, \bibinfo{title}{Density functional perturbation
  theory for gated two-dimensional heterostructures: Theoretical developments
  and application to flexural phonons in graphene}, \bibinfo{journal}{Phys.
  Rev. B} \bibinfo{volume}{96} (\bibinfo{year}{2017}{\natexlab{b}})
  \bibinfo{pages}{075448}.

\bibitem[{per(????)}]{perturbo}
\bibinfo{title}{The {\sc Perturbo} website},
  \bibinfo{note}{\url{https://perturbo-code.github.io}}

\bibitem[{Dal~Corso(2001)}]{DalCorso2001}
\bibinfo{author}{A.~Dal~Corso}, \bibinfo{title}{Density-functional perturbation
  theory with ultrasoft pseudopotentials}, \bibinfo{journal}{Phys. Rev. B}
  \bibinfo{volume}{64}~(\bibinfo{number}{23}) (\bibinfo{year}{2001})
  \bibinfo{pages}{235118}.

\bibitem[{Maradudin and Vosko(1968)}]{MARADUDIN1968}
\bibinfo{author}{A.~A. Maradudin}, \bibinfo{author}{S.~H. Vosko},
  \bibinfo{title}{Symmetry properties of the normal vibrations of a crystal},
  \bibinfo{journal}{Rev. Mod. Phys.} \bibinfo{volume}{40}~(\bibinfo{number}{1})
  (\bibinfo{year}{1968}) \bibinfo{pages}{1--37}.

\bibitem[{Poncé et~al.(2016)Poncé, Margine, Verdi, and Giustino}]{Ponce2016}
\bibinfo{author}{S.~Poncé}, \bibinfo{author}{E.~R. Margine},
  \bibinfo{author}{C.~Verdi}, \bibinfo{author}{F.~Giustino},
  \bibinfo{title}{{EPW}: {Electron}–phonon coupling, transport and
  superconducting properties using maximally localized {Wannier} functions},
  \bibinfo{journal}{Comput. Phys. Commun.} \bibinfo{volume}{209}
  (\bibinfo{year}{2016}) \bibinfo{pages}{116--133}.

\bibitem[{Gaddemane et~al.(2019)Gaddemane, Gopalan, Van~de Put, and
  Fischetti}]{Gaddemane2019}
\bibinfo{author}{G.~Gaddemane}, \bibinfo{author}{S.~Gopalan},
  \bibinfo{author}{M.~Van~de Put}, \bibinfo{author}{M.~V. Fischetti},
  \bibinfo{title}{Limitations of ab initio methods to predict the
  electronic-transport properties of two-dimensional materials: {The}
  computational example of {2H}-phase transition metal dichalcogenides},
  \bibinfo{journal}{arXiv:1912.03795 [cond-mat.mtrl-sci]} .

\bibitem[{van Setten et~al.(2018)van Setten, Giantomassi, Bousquet, Verstraete,
  Hamann et~al.}]{Setten2018}
\bibinfo{author}{M.~J. van Setten}, \bibinfo{author}{M.~Giantomassi},
  \bibinfo{author}{E.~Bousquet}, \bibinfo{author}{M.~J. Verstraete},
  \bibinfo{author}{D.~R. Hamann}, et~al., \bibinfo{title}{The {PseudoDojo}:
  {Training} and grading a 85 element optimized norm-conserving pseudopotential
  table}, \bibinfo{journal}{Comput. Phys. Commun.} \bibinfo{volume}{226}
  (\bibinfo{year}{2018}) \bibinfo{pages}{39--54}.

\bibitem[{Damle et~al.(2017)Damle, Lin, and Ying}]{Damle2017}
\bibinfo{author}{A.~Damle}, \bibinfo{author}{L.~Lin},
  \bibinfo{author}{L.~Ying}, \bibinfo{title}{{SCDM}-k: {Localized} orbitals for
  solids via selected columns of the density matrix}, \bibinfo{journal}{J.
  Comput. Phys.} \bibinfo{volume}{334} (\bibinfo{year}{2017})
  \bibinfo{pages}{1--15}.

\bibitem[{Blood(1972)}]{Blood1972}
\bibinfo{author}{P.~Blood}, \bibinfo{title}{Electrical properties of $n$-type
  epitaxial {GaAs} at high temperatures}, \bibinfo{journal}{Phys. Rev. B}
  \bibinfo{volume}{6}~(\bibinfo{number}{6}) (\bibinfo{year}{1972})
  \bibinfo{pages}{2257--2261}.

\bibitem[{Nichols et~al.(1980)Nichols, Yee, and Wolfe}]{Nichols1980}
\bibinfo{author}{K.~H. Nichols}, \bibinfo{author}{C.~M.~L. Yee},
  \bibinfo{author}{C.~M. Wolfe}, \bibinfo{title}{High-temperature carrier
  transport in $n$-type epitaxial {GaAs}}, \bibinfo{journal}{Solid-State
  Electron.} \bibinfo{volume}{23}~(\bibinfo{number}{2}) (\bibinfo{year}{1980})
  \bibinfo{pages}{109--116}.

\bibitem[{Lee et~al.(2019)Lee, Zhou, Chen, and Bernardi}]{Lee2019}
\bibinfo{author}{N.-E. Lee}, \bibinfo{author}{J.-J. Zhou},
  \bibinfo{author}{H.-Y. Chen}, \bibinfo{author}{M.~Bernardi},
  \bibinfo{title}{Next-to-{leading} {order} {ab} {initio} {electron}-{phonon}
  {scattering}}, \bibinfo{journal}{arXiv:1903.08261 [cond-mat.mes-hall]} .

\bibitem[{Ma et~al.(2018)Ma, Nissimagoudar, and Li}]{Ma2018}
\bibinfo{author}{J.~Ma}, \bibinfo{author}{A.~S. Nissimagoudar},
  \bibinfo{author}{W.~Li}, \bibinfo{title}{First-principles study of electron
  and hole mobilities of {Si} and {GaAs}}, \bibinfo{journal}{Phys. Rev. B}
  \bibinfo{volume}{97}~(\bibinfo{number}{4}) (\bibinfo{year}{2018})
  \bibinfo{pages}{045201}.

\bibitem[{﻿Madelung et~al.(2002)﻿Madelung, R{\"o}ssler, and
  Schulz}]{Madelung2002}
\bibinfo{editor}{O.~﻿Madelung}, \bibinfo{editor}{U.~R{\"o}ssler},
  \bibinfo{editor}{M.~Schulz} (Eds.), \bibinfo{title}{Semiconductors}, vol.
  \bibinfo{volume}{﻿41A1$\beta$}, \bibinfo{publisher}{Springer, Berlin},
  \bibinfo{year}{2002}.

\bibitem[{Homm et~al.(2008)Homm, Klar, Teubert, and Heimbrodt}]{Homm2008}
\bibinfo{author}{G.~Homm}, \bibinfo{author}{P.~J. Klar},
  \bibinfo{author}{J.~Teubert}, \bibinfo{author}{W.~Heimbrodt},
  \bibinfo{title}{Seebeck coefficients of n-type (Ga,In)(N,As), (B,Ga,In)As,
  and GaAs}, \bibinfo{journal}{Appl. Phys. Lett.}
  \bibinfo{volume}{93}~(\bibinfo{number}{4}) (\bibinfo{year}{2008})
  \bibinfo{pages}{042107}.

\bibitem[{Protik and Broido(2019)}]{Protik2019}
\bibinfo{author}{N.~H. Protik}, \bibinfo{author}{D.~A. Broido},
  \bibinfo{title}{Coupled transport of phonons and carriers in semiconductors},
  \bibinfo{journal}{arXiv:1911.02916 [cond-mat.mtrl-sci]} .

\bibitem[{Yu et~al.(2016)Yu, Ong, Pan, Cui, Xin, Shi et~al.}]{Yu2016}
\bibinfo{author}{Z.~Yu}, \bibinfo{author}{Z.-Y. Ong}, \bibinfo{author}{Y.~Pan},
  \bibinfo{author}{Y.~Cui}, \bibinfo{author}{R.~Xin}, \bibinfo{author}{Y.~Shi},
  et~al., \bibinfo{title}{Realization of room-temperature phonon-limited
  carrier transport in monolayer MoS$_{2}$ by dielectric and carrier
  screening}, \bibinfo{journal}{Adv. Mater.}
  \bibinfo{volume}{28}~(\bibinfo{number}{3}) (\bibinfo{year}{2016})
  \bibinfo{pages}{547--552}.

\bibitem[{Sadasivam et~al.(2017)Sadasivam, Chan, and Darancet}]{Sadasivam2017}
\bibinfo{author}{S.~Sadasivam}, \bibinfo{author}{M.~K.~Y. Chan},
  \bibinfo{author}{P.~Darancet}, \bibinfo{title}{Theory of thermal relaxation
  of electrons in semiconductors}, \bibinfo{journal}{Phys. Rev. Lett.}
  \bibinfo{volume}{119} (\bibinfo{year}{2017}) \bibinfo{pages}{136602}.

\bibitem[{Molina-Sánchez et~al.(2017)Molina-Sánchez, Sangalli, Wirtz, and
  Marini}]{Molina2017}
\bibinfo{author}{A.~Molina-Sánchez}, \bibinfo{author}{D.~Sangalli},
  \bibinfo{author}{L.~Wirtz}, \bibinfo{author}{A.~Marini}, \bibinfo{title}{Ab
  initio calculations of ultrashort carrier dynamics in two-dimensional
  materials: Valley depolarization in single-layer WSe$_{2}$},
  \bibinfo{journal}{Nano Lett.} \bibinfo{volume}{17}~(\bibinfo{number}{8})
  (\bibinfo{year}{2017}) \bibinfo{pages}{4549--4555}.

\bibitem[{Zhou and Bernardi(2019)}]{Zhou2019}
\bibinfo{author}{J.-J. Zhou}, \bibinfo{author}{M.~Bernardi},
  \bibinfo{title}{Predicting charge transport in the presence of polarons: The
  beyond-quasiparticle regime in ${\mathrm{SrTiO}}_{3}$},
  \bibinfo{journal}{Phys. Rev. Research} \bibinfo{volume}{1}
  (\bibinfo{year}{2019}) \bibinfo{pages}{033138}.

\bibitem[{Hellman et~al.(2013)Hellman, Steneteg, Abrikosov, and
  Simak}]{Hellman2013}
\bibinfo{author}{O.~Hellman}, \bibinfo{author}{P.~Steneteg},
  \bibinfo{author}{I.~A. Abrikosov}, \bibinfo{author}{S.~I. Simak},
  \bibinfo{title}{Temperature dependent effective potential method for accurate
  free energy calculations of solids}, \bibinfo{journal}{Phys. Rev. B}
  \bibinfo{volume}{87}~(\bibinfo{number}{10}) (\bibinfo{year}{2013})
  \bibinfo{pages}{104111}.

\end{thebibliography}
\bibliographystyle{elsarticle-num-names}
\end{document}